\documentclass[a4paper, 11pt]{article}
\usepackage[margin = 2.5cm]{geometry}

\usepackage[colorlinks, allcolors = blue]{hyperref}
\usepackage{amsmath}
\usepackage{amssymb}
\usepackage{mathtools}
\usepackage{slashed}

\numberwithin{equation}{section}

\def \bR {\mathbb{R}}

\newcommand{\bea}{\begin{eqnarray}}
\newcommand{\eea}{\end{eqnarray}}

\def \defeq {\vcentcolon=} % DEFINED EQUAL TO
\def \dd {d} % DIFFERENTIAL, d
\def \hodge #1{\,{}^\star\! #1} % HODGE DUAL
\def \e {\mathbf{e}} % FRAME
\def \H #1#2{\left \langle #1, #2 \right \rangle} % HERMITIAN INNER PRODUCT

\begin{document}

\begin{center}
    \vspace*{-1.0 cm}
    \vspace{2.0 cm}{
        \Large \bf TCFHs and hidden symmetries of type IIA AdS backgrounds
    } \\[0.2cm]

    \vskip 2.0cm
        G.~ Papadopoulos and J.~Phillips \\
    \vskip 0.6cm

    \begin{small}
        \textit{
            Department of Mathematics \\
            King's College London \\
            Strand \\
            London WC2R 2LS, UK
        } \\
		\texttt{george.papadopoulos@kcl.ac.uk} \\
        \texttt{jake.phillips@kcl.ac.uk}
    \end{small}
    \\*[0.6cm]
\end{center}

\vskip 2.5 cm

\begin{abstract}
    \noindent
    We present the twisted covariant form hierarchies (TCFHs)  of warped (massive) IIA AdS backgrounds. As a consequence we demonstrate that all Killing spinor form bilinears satisfy a generalisation of the conformal Killing-Yano equation with respect to the TCFH connections. We also explore some of the properties of TCFHs which include  the reduced  holonomy  of the  minimal TCFH connections for generic backgrounds. Furthermore, we investigate the interplay between TCFHs and hidden   symmetries of  probes propagating on IIA AdS backgrounds.  We find that some of the Killing spinor form bilinears of near horizon geometries of a class of IIA intersecting brane configurations are Killing-Yano forms  and so generate hidden symmetries for spinning particle probes.
\end{abstract}

\vskip 1.5 cm

\newpage

\section{Introduction}

Recently it has been demonstrated  in \cite{gptcfh}, following earlier work in \cite{jggp}, that the conditions induced by the gravitino Killing spinor equation (KSE) on the (Killing spinor) form bilinears of any supergravity theory, which may include higher curvature corrections, can be organised as a TCFH.  This means that there is a connection ${\cal D}^{\cal F}$ which depends on the fluxes, ${\cal F}$, of the theory such that
\bea
{\cal D}_X^{\cal F}\Omega=i_X {\cal P}+X\wedge {\cal Q}~,
\label{tcfhcon}
\eea
for any vector field $X$ on the spacetime, where $\Omega$ is spanned by the form bilinears and ${\cal P}$ and ${\cal Q}$ are multiforms which depend on $\Omega$ and ${\cal F}$. The TCFH connection ${\cal D}^{\cal F}$ may not be form degree preserving. A consequence of (\ref{tcfhcon}) is that $\Omega$ satisfies a generalisation of the conformal Killing-Yano (CKY) equation\footnote{The standard CKY equation reads $\nabla_X\omega=i_X d\omega-{1\over n-k+1} X\wedge\delta\omega$, where $\nabla$ is the Levi-Civita connection and $\omega$ a $k$-form. If $\omega$ is co-closed, $\delta\omega=0$, then $\omega$ is a KY form. If $\omega$ is closed, $d\omega=0$, then $\omega$ is a closed CKY (CCKY) form. The Hodge dual of a KY form is a CCKY form and vice-versa.} with respect to ${\cal D}^{\cal F}$. Killing-Yano (KY) forms have played a crucial role  in the integrability of geodesic flows of several black hole spacetimes, beginning with the  Kerr black hole in \cite{carter-b, penrose, floyd},  as well as other classical field equations on curved backgrounds; for some selected publications see \cite{chandrasekhar, carter-a, carter-c, page, sfetsos, lun} and the reviews \cite{revky, frolov}.  For additional applications of CKY, KY and CCKY forms see e.g. \cite{kt, gpky, hl, ls1, ls2}.
 Moreover, it has been demonstrated in \cite{gibbons} that spinning particle probes \cite{bvh} propagating on backgrounds equipped  with a KY form admit (hidden) symmetries generated by the form.   This raises the possibility that, as a consequence of TCFH, the form bilinears of supersymmetric backgrounds may be associated with the (hidden) symmetries of certain probes whose actions may include couplings associated with the supergravity fields. Thus, there may be an interplay between TCFHs and probe conservation laws.

The construction of the TCFH for 11-dimensional, IIA and IIB supergravities on generic supersymmetric backgrounds can be found in \cite{epbgp1, lggpjp}.  Similar results have been obtained in some lower dimensional supergravity theories \cite{epbgp2}. In all cases, it has been demonstrated that there are supersymmetric backgrounds whose form bilinears generate symmetries for suitably chosen probe actions, i.e. it has been found  that the invariance conditions of the probe actions  match those  associated with the TCFH on the form bilinears.   Moreover the TCFHs of all 11-dimensional and IIB  AdS backgrounds have been presented in \cite{epbgp, lggp}. An investigation of the relation between TCFHs and invariance conditions for probes has also been presented for AdS backgrounds yielding similar results.

The purpose of this paper is to present the TCFH on the internal spaces of all warped AdS backgrounds of (massive) IIA supergravity \cite{romans}. In addition some of their properties are explored which include the reduced holonomy of the minimal connection for generic supersymmetric backgrounds. Next we investigate the question on whether some of the form bilinears generate symmetries for spinning particles propagating on such backgrounds. It is demonstrated that this is the case for a class of AdS backgrounds constructed using ansatze that include the near horizon geometries of some IIA intersecting brane configurations. This work completes the construction of TCFHs for all AdS backgrounds of type II supergravities in 10- and 11-dimensions.

This paper has been organised as follows.  In sections 2, 3 and 4, the TCFH of warped IIA AdS$_k$, $k=2,3,4$ backgrounds are presented. This includes also the investigation of some of the properties of the TCFH connections, such as their holonomy. In section 5, the TCFH of warped IIA AdS$_k$, $k=5,6,7$ backgrounds
are given. In section 6, we present some explicit examples where the TCFH generates symmetries for spinning particles propagating on the internal space of AdS$_2$ and AdS$_3$ backgrounds, and in section 7 we give our conclusions.

\section{The TCFH of warped AdS\textsubscript{2} backgrounds}

The approach that we shall follow below to construct the TCFHs on the internal spaces of  all  warped AdS backgrounds of massive IIA supergravity is based on the solution of the KSEs of the theory presented in \cite{ggkpiia, bgpiia}. In these works  the KSEs of the theory are integrated over the AdS subspace of warped AdS backgrounds without any additional assumptions on the form of the Killing spinors.  Then the remaining independent KSEs on the internal space of the AdS backgrounds are identified. A similar procedure is used for the field equations of the theory. The main advantage of this method is that it does not involve additional assumptions, such as a certain factorisation  of Killing spinors, and so it is general. For a comparison of the different methods to solve the  KSEs of warped AdS backgrounds see \cite{adsdes}.

\subsection{Fields and Killing spinors}

Let $\Phi$ be the dilaton, and $G$, $H$, $F$ be the 4-, 3- and 2-form field strengths of (massive) IIA supergravity, respectively. The bosonic fields of a warped AdS$_2$ background, AdS$_2 \times_w M^8$, with internal space $M^8$  can be expressed as follows
\bea
		&&g = 2\, \e^+ \e^- + g(M^8)~,~~~
		G = \e^+ \wedge \e^- \wedge X + Y~,~~~ H = \e^+ \wedge \e^- \wedge W + Z~,~~~
\cr
&&
		F = N \, \e^+ \wedge \e^- + P\,,~~~ \quad S = m e^\Phi\,,~~~ \quad \Phi = \Phi~,
	\eea
where now the dilaton field $\Phi \in C^\infty(M^8)$, $g(M^8)$ is a metric on the internal space $M^8$, and $N \in C^\infty(M^8)$, $W \in \Omega^1(M^8)$, $X, P \in \Omega^2(M^8)$, $Z \in \Omega^3(M^8)$ and $Y \in \Omega^4(M^8)$. Moreover $m$ is a constant that is non-zero in massive IIA and vanishes in standard IIA supergravity. We have also introduced the pseudo-orthonormal (co-)frame
\begin{equation}
		\e^+ = \dd u~, \quad \e^- = \dd r -2 r A^{-1} dA - \frac{1}{2} r^2 \ell^{-2} A^{-2} \dd u~, \quad \e^i = e^i{}_J \dd y^J~,
\label{ads2frame}
\end{equation}
on AdS$_2 \times_w M^8$, where $A\in C^\infty(M^8)$ is the warp factor, $\e^i$ is an orthonormal frame on $M^8$  that depends only on the coordinates $y$ of $M^8$,  $g(M^8)=\delta_{ij} \e^i \e^j$, and $\ell$ is the radius of AdS\textsubscript{2}.  Moreover $(u, r)$ are the remaining  coordinates of the spacetime. It can be seen after a coordinate transformation that the spacetime metric $g$ can be put into the standard warped form $g=A^2 g_{\ell}(AdS_2)+ g(M^8)$, where
$g_{\ell}(AdS_2)$ is the standard metric on AdS$_2$ with radius $\ell$.

The KSEs of massive IIA supergravity for warped AdS$_2$ backgrounds have been integrated over the $(u,r)$ coordinates  in  \cite{ggkpiia, bgpiia}. In such a case, the Killing spinors can be expressed as $\epsilon=\epsilon(u,r, \eta_\pm)$, where $\eta_\pm$ are spinors that depend only on the coordinates of $M^8$
and satisfy $\Gamma_\pm\eta_\pm=0$, where the gamma matrices $(\Gamma_+, \Gamma_-, \Gamma_i)$ are taken with respect to the frame (\ref{ads2frame}). The precise expression for $\epsilon$ in terms of $u,r$ and $\eta_\pm$, which can be found in \cite{bgpiia}, is not essential in what follows and so it will not be presented here.
Furthermore, the conditions that gravitino KSE imposes on  $\eta_\pm$ along  $M^8$ are
\begin{equation} \label{AdS2KSE}
	\mathcal{D}_m^{(\pm)} \eta_\pm = 0~,
\end{equation}
   where
\begin{equation} \label{AdS2_Connection}
	\begin{split}
			\mathcal{D}_m^{(\pm)}\eta_\pm = \nabla_m \eta_\pm\, \pm & \frac{1}{2} A^{-1} \partial_{m} A \,\eta_\pm \mp \frac{1}{16} \slashed{X} \Gamma_{m} \eta_\pm + \frac{1}{8 \cdot 4 !} \slashed{Y} \Gamma_{m} \eta_\pm + \frac{1}{8} S \Gamma_{m} \eta_\pm \\
		&+\Gamma_{11}\left(\mp \frac{1}{4} W_{m} \eta_\pm + \frac{1}{8} \slashed{Z}_{m} \eta_\pm \pm \frac{1}{8} N \Gamma_{m} \eta_\pm - \frac{1}{16} \slashed{P} \Gamma_{m} \eta_\pm \right)~,
	\end{split}
\end{equation}
is the supercovariant connection\footnote{We use the conventions of \cite{ggkpiia, bgpiia}.  In particular if $\alpha$ is a $k$-form on $M^8$, then $\slashed{\alpha}=\alpha_{j_1\dots j_k} \Gamma^{j_i\dots j_k}$ and $\slashed \alpha_i=\alpha_{i j_1\dots j_{k-1}} \Gamma^{j_i\dots j_{k-1}}$.} on $M^8$, $m=1,\dots,8$
and  $\nabla$ is the spin  connection associated with the metric $g(M^8)$. These are clearly parallel transport equations for $\eta_\pm$.  The Killing spinors $\eta_\pm$ satisfy additional conditions \cite{bgpiia} arising from the dilatino KSE of massive IIA supergravity. But these additional conditions are not essential for the
TCFH below and so we shall not describe them here. However, they will be used later when we discuss examples and some aspects of them will be summarised there.

\subsection{The TCFH on $M^8$} \label{sec:TCFH}

It has been demonstrated in \cite{gptcfh} that the conditions imposed on the Killing spinor bilinears by the gravitino KSE of any supergravity theory
can be organised as a TCFH. Here we shall focus on the TCFH associated with the form bilinears on $M^8$ constructed from the Killing spinors $\eta_\pm$ satisfying the KSEs (\ref{AdS2KSE}).  Given two such Killing spinors $\eta_\pm^r$ and $\eta^s_\pm$, one can define the $k$-form bilinears
\begin{equation}
	\phi_\pm^{rs} = \frac{1}{k!} \H{\eta_\pm^r}{\Gamma_{i_1 \dots i_k} \eta_\pm^s}\, \e^{i_1} \wedge \dots \wedge \e^{i_k}~,\qquad \tilde\phi_\pm^{rs} = \frac{1}{k!} \H{\eta_\pm^r}{\Gamma_{i_1 \dots i_k} \Gamma_{11} \eta_\pm^s}\, \e^{i_1} \wedge \dots \wedge \e^{i_k}~,
\label{fbi}
\end{equation}
where $\H{\cdot}{\cdot}$ denotes the spin-invariant  inner product on $M^8$ for which the spacelike gamma matrices are Hermitian while the time-like ones are anti-Hermitian.

Because of the reality condition on $\eta_\pm$, which follows from that of IIA Killing spinors, the form bilinears are either symmetric or skew-symmetric on the exchange of  $\eta^r$ and $\eta^s$. A basis in the space of form bilinears\footnote{Note that the form bilinears constructed from $\eta_+$ and $\eta_-$ spinors vanish.}  on $M^8$, up to Hodge duality\footnote{Our Hodge duality conventions are $\hodge{\omega}_{m_1 \dots m_{n-p}} = \frac{1}{p!}\omega_{q_1\dots q_p}\epsilon^{q_1\dots q_p}{}_{m_1\dots m_{n-p}}$, where $\omega$ is a $p$-form on a $n$-dimensional Riemannian manifold $M^n$ with orientation chosen as $\epsilon_{12\dots n}=1$.}, which are symmetric in the exchange of Killing spinors is
\begin{equation} \label{AdS2_Sym_Bilinears}
	\begin{gathered}
		f^{rs}_\pm = \H{\eta^r_\pm}{\eta^s_\pm} , \quad \tilde{f}^{rs}_\pm = \H{\eta^r_\pm}{\Gamma_{11} \eta^s_\pm} , \quad k^{rs}_\pm = \H{\eta^r_\pm}{\Gamma_ i\eta^s_\pm} \, \e^i~, \\
		\tilde{\pi}^{rs}_\pm = \frac{1}{3!} \H{\eta^r_\pm}{\Gamma_ {ijk}\Gamma_{11} \eta^s_\pm} \, \e^i \wedge \e^j \wedge \e^k, \quad \zeta^{rs}_\pm = \frac{1}{4!} \H{\eta^r_\pm}{\Gamma_{i_1 \dots i_4}\eta^s_\pm}\, \e^{i_1} \wedge \dots \wedge \e^{i_4}~.
	\end{gathered}
\end{equation}

To find the TCFH associated to the above form bilinears note  that
\begin{equation}
	\nabla_m \phi_\pm{}^{rs}_{i_1 \dots i_k} = \H{\nabla_m \eta_\pm^r}{\Gamma_{i_1 \dots i_k} \eta_\pm^s} + \H{\eta_\pm^r}{\Gamma_{i_1 \dots i_k} \nabla_m \eta_\pm^s}~,
\end{equation}
and similarly for $\tilde \phi_\pm{}^{rs}$.  Then using the KSEs \eqref{AdS2KSE}, one can replace in the right-hand-side of the above equation the derivatives on the spinors in term of a Clifford algebra element constructed from the fluxes of the theory. After some extensive Clifford algebra computation, one can demonstrate that the right-hand-side can always be
organised as a TCFH.

In particular,  the TCFH of the form bilinears (\ref{AdS2_Sym_Bilinears}), with respect to the minimal connection\footnote{See \cite{gptcfh} for the definition.}  $\mathcal{D}^\mathcal{F}$ is
\begin{equation}
    \begin{split}
	    {\cal D}^{\cal F}_m f_\pm \defeq& \nabla_m f_\pm \\
        =& \mp A^{-1} \partial_{m} A \, f_\pm \mp \frac{1}{4} X_{mp}k_\pm{}^p \pm \frac{1}{4!} \hodge{Y}_{mpqr} \tilde{\pi}_\pm{}^{pqr} \\
		&- \frac{1}{4}Sk_\pm{}_m \pm\frac{1}{2}W_m\tilde{f}_\pm-\frac{1}{8}P_{pq}\tilde{\pi}_\pm{}^{pq}{}_m~,
    \end{split}
\end{equation}

\begin{equation}
    \begin{split}
	    {\cal D}^{\cal F}_m \tilde{f}_\pm \defeq& \nabla_m \tilde{f}_\pm \\
        =& \mp A^{-1} \partial_{m} A \, \tilde{f}_\pm \mp \frac{1}{8} X_{pq}\tilde{\pi}_\pm{}^{pq}{}_m - \frac{1}{4!} Y_{mpqr} \tilde{\pi}_\pm{}^{pqr} \\
		&\pm \frac{1}{2} W_m f_\pm \mp \frac{1}{4}Nk_\pm {}_m - \frac{1}{4} P_{mp}k_\pm{}^p~,
    \end{split}
\end{equation}

\begin{equation}
    \begin{split}
		{\cal D}^{\cal F}_m k_\pm{}_i \defeq& \nabla_m k_\pm{}_i + \frac{1}{12}Y_{mpqr}\zeta_\pm{}^{pqr}{}_i + \frac{1}{4}Z_{mpq}\tilde{\pi}_\pm{}^{pq}{}_i \\
        =& \mp A^{-1} \partial_{m} A \, k_\pm{}_i \, \mp \frac{1}{8}X_{pq}\zeta_\pm{}^{pq}{}_{mi} \mp \frac{1}{4} X_{mi} f_\pm - \frac{1}{4 \cdot 4!} \delta_{mi}Y_{p_1 \dots p_4} \zeta_\pm{}^{p_1 \dots p_4}  \\
		&+ \frac{1}{12}Y_{[m|pqr|}\zeta_\pm {}^{pqr}{}_{i]} - \frac{1}{4}\delta_{mi} Sf_\pm \pm \frac{1}{4}\delta_{mi} N \tilde{f}_\pm \mp \frac{1}{4 \cdot 4!} \hodge{P}_{mip_1\dots p_4}\zeta_\pm{}^{p_1 \dots p_4} \\
		&+ \frac{1}{4} P_{mi}\tilde{f}_\pm~,
    \end{split}
\end{equation}

\begin{equation}
	\begin{split}
		{\cal D}^{\cal F}_m \tilde{\pi}_\pm{}_{ijk} \defeq& \nabla_m \tilde{\pi}_\pm{}_{ijk} + \frac{1}{4} \hodge{X}_{m[ij|pqr|}\zeta_\pm{}^{pqr}{}_{k]} \pm \frac{3}{4} \hodge{Y}_{m[i|pq|}\zeta_\pm{}^{pq}{}_{jk]} \pm \frac{3}{4}\hodge{Z}_{m[ij|pq|}\tilde\pi_\pm{}^{pq}{}_{k]} \\
		&- \frac{3}{2}Z_{m[ij}k_\pm{}_{k]} - \frac{1}{2}P_{mp}\zeta_\pm{}^p{}_{ijk} \\
		=& \mp A^{-1} \partial_{m} A \, \tilde{\pi}_\pm{}_{ijk} \, \pm \frac{3}{4}\delta_{m[i}X_{jk]}\tilde{f}_\pm -\frac{1}{32} \delta_{m[i}\hodge{X}_{jk]p_1 \dots p_4}\zeta_\pm{}^{p_1 \dots p_4} \\
		&+ \frac{1}{6} \hodge{X}_{[mij|pqr|}\zeta_\pm{}^{pqr}{}_{k]} \pm \frac{1}{4}\hodge{Y}_{mijk} f_\pm + \frac{1}{4} Y_{mijk}\tilde{f}_\pm \pm \frac{1}{4}\delta_{m[i}\hodge{Y}_{j|pqr|}\zeta_\pm {}^{pqr}{}_{k]} \\
		& \pm \frac{3}{4}\hodge{Y}_{[mi|pq|}\zeta_\pm{}^{pq}{}_{jk]} \pm\frac{1}{4 \cdot 4!} \hodge{S}_{mijkp_1\dots p_4}\zeta_\pm {}^{p_1\dots p_4} \pm \frac{1}{4} \delta_{m[i}\hodge{Z}_{jk]pqr} \tilde\pi_\pm {}^{pqr} \\
		& \pm \hodge{Z}_{[mij|pqr|}\tilde\pi_\pm{}^{pq}{}_{k]} \pm \frac{1}{4} N\zeta_\pm{}_{mijk} + \frac{3}{8} \delta_{m[i|}P_{pq|}\zeta_\pm{}^{pq}{}_{jk]} - P_{[m|p|}\zeta_\pm{}^p{}_{ijk]} \\
		&- \frac{3}{4}\delta_{m[i}P_{jk]}f_\pm~,
	\end{split}
\end{equation}

\begin{equation}
	\begin{split}
		{\cal D}^{\cal F}_m \zeta_\pm{}_{i_1 \dots i_4} \defeq& \nabla_m \zeta_\pm{}_{i_1 \dots i_4} -\hodge{X}_{m[i_1i_2i_3|pq|}\tilde\pi_\pm{}^{pq}{}_{i_4]} - 2 Y_{m[i_1i_2i_3}k_\pm{}_{i_4]} \pm \,3 \hodge{Y}_{m[i_1i_2|p|}\tilde\pi_\pm{}^p{}_{i_3i_4]} \\
		&+ \frac{1}{2 \cdot 4!}W_m \epsilon_{i_1 \dots i_4}{}^{j_1 \dots j_4}\zeta_\pm{}_{j_1 \dots j_4} \pm \frac{3}{2} \hodge{Z}_{m[i_1i_2|pq|}\zeta_\pm{}^{pq}{}_{i_3i_4]} + 2 P_{m[i_1}\tilde\pi_\pm{}_{i_2i_3i_4]} \\
		=& \mp A^{-1} \partial_{m} A \, \zeta_\pm{}_{i_1 \dots i_4} \, \pm 3\delta_{m[i_1}X_{i_2i_3}k_\pm{}_{i_4]} -\frac{1}{6}\delta_{m[i_1}\hodge{X}_{i_2i_3i_4]pqr}\tilde\pi_\pm{}^{pqr} \\
		&- \frac{5}{8}\hodge{X}_{[mi_1i_2i_3|pq|}\tilde\pi_\pm{}^{pq}{}_{i_4]} - \delta_{m[i_1}Y_{i_2i_3i_4]p}k_\pm{}^p - \frac{5}{4}Y_{[mi_1i_2i_3} k_\pm{}_{i_4]} \\
		&\pm \frac{5}{2} \hodge{Y}_{[mi_1i_2|p|}\tilde\pi_\pm{}^p{}_{i_3i_4]}  \mp \frac{3}{2}\delta_{m[i_1}\hodge{Y}_{i_2i_3|pq|}\tilde\pi_\pm{}^{pq}{}_{i_4]} \pm \frac{1}{24} \hodge{S}_{mi_1\dots i_4 pqr}\tilde\pi_\pm{}^{pqr}  \\
		&\pm \delta_{m[i_1}\hodge{Z}_{i_2 i_3|pqr|}\zeta_\pm{}^{pqr}{}_{i_4]} \pm \frac{5}{2}\hodge{Z}_{[mi_1i_2|pq|}\zeta_\pm{}^{pq}{}_{i_3i_4]} \mp N\delta_{m[i_1}\tilde\pi_\pm{}_{i_2i_3i_4]} \\
		&\mp \frac{1}{4} \hodge{P}_{mi_1\dots i_4 p}k_\pm{}^p + 3\delta_{m[i_1}P_{i_2|p|}\tilde\pi_\pm{}^p{}_{i_3i_4]} + \frac{5}{2}P_{[mi_1}\tilde\pi_\pm{}_{i_2i_3i_4]}~,
	\end{split}
\end{equation}
where for simplicity we have suppressed the $r,s$ indices on the form bilinears that label the different Killing spinors. It is clear that the above conditions on the form bilinears are of the form of a TCFH as in (\ref{tcfhcon}).

A basis in the space of form bilinears on $ M^8$, up to Hodge duality, which are skew-symmetric in the exchange of  $\eta^r$ and $\eta^s$ is the following
\begin{equation} \label{AdS2_Skew_Bilinears}
	\begin{gathered}
		\tilde{k}^{rs}_{\pm} = \H{\eta^r_\pm}{\Gamma_i \Gamma_{11} \eta^s_\pm} \, \e^i~, \quad \omega^{rs}_\pm = \frac{1}{2} \H{\eta^r_\pm}{\Gamma_{ij}\eta^s_\pm} \, \e^i \wedge \e^j~, \\
		\tilde{\omega}^{rs}_\pm = \frac{1}{2} \H{\eta^r_\pm}{\Gamma_{ij}\Gamma_{11}\eta^s_\pm} \, \e^i \wedge \e^j~, \quad	\pi^{rs}_\pm = \frac{1}{3!} \H{\eta^r_\pm}{\Gamma_ {ijk}\eta^s_\pm} \, \e^i \wedge \e^j \wedge \e^k~.
	\end{gathered}
\end{equation}
The associated TCFH with respect to the minimal connection, $\mathcal{D}^\mathcal{F}$, is given by
\begin{equation}
	\begin{split}
		{\cal D}^{\cal F}_m \tilde k_\pm{}_i \defeq& \nabla_m \tilde k_\pm{}_i \pm \frac{1}{2} X_{mp} \tilde\omega_\pm{}^p{}_i + \frac{1}{4} Z_{mpq} \pi_\pm{}^{pq}{}_i - \frac{1}{2} P_{mp} \omega_\pm{}^p{}_i \\
		=& \mp A^{-1} \partial_{m} A \, \tilde{k}_\pm{}_i \, \mp \frac{1}{8} \delta_{mi} X_{pq} \tilde\omega_\pm{}^{pq} \pm \frac{1}{2} X_{[m|p|}\tilde\omega_\pm{}^p{}_{i]} \\
		&  \mp \frac{1}{8} \hodge{Y}_{mipq} \omega_\pm {}^{pq} -\frac{1}{8} Y_{mipq} \tilde\omega_\pm{}^{pq} - \frac{1}{4} S \tilde\omega_\pm{}_{mi} \\
		& \pm \frac{1}{4} N \omega_\pm{}_{mi} + \frac{1}{8} \delta_{mi}P_{pq} \omega_\pm{}^{pq} - \frac{1}{2} P_{[m|p|}\omega_\pm{}^p{}_{i]}~,
	\end{split}
\end{equation}

\begin{equation}
	\begin{split}
		{\cal D}^{\cal F}_m \omega_\pm{}_{ij} \defeq& \nabla_m \omega_\pm{}_{ij} \pm \frac{1}{2} X_{mp} \pi_\pm{}^p{}_{ij} + \frac{1}{2} Y_{m[i|pq|}\pi_\pm{}^{pq}{}_{j]} \mp \frac{1}{2} W_m \tilde\omega_{\pm ij} \\
		&+ Z_{m[i|p|}\tilde\omega_\pm{}^p{}_{j]} + P_{m[i} \tilde k_{\pm j]} \\
		=& \mp A^{-1} \partial_{m} A \, \omega_\pm{}_{ij} \, \mp \frac{1}{4} \delta_{m[i}X_{|pq|}\pi_\pm{}^{pq}{}_{j]} \pm \frac{3}{4} X_{[m|p|}\pi_\pm{}^p{}_{ij]} \\
		& \mp \frac{1}{4} \hodge{Y}_{mijp} \tilde k_\pm{}^p + \frac{1}{12} \delta_{m[i}Y_{j]pqr}\pi_\pm{}^{pqr} + \frac{3}{8} Y_{[mi|pq|}\pi_\pm{}^{pq}{}_{j]} \\
		&- \frac{1}{4} S \pi_{\pm mij} \mp \frac{1}{2} N \delta_{m[i}\tilde k_{\pm j]} \pm \frac{1}{4!} \hodge{P}_{mijpqr} \pi_\pm{}^{pqr} \\
		&+ \frac{1}{2} \delta_{m[i}P_{j]p}\tilde k_\pm{}^p + \frac{3}{4} P_{[mi} \tilde k_{\pm j]}~,
	\end{split}
\end{equation}

\begin{equation}
	\begin{split}
		{\cal D}^{\cal F}_m \tilde\omega_\pm{}_{ij} \defeq& \nabla_m \tilde\omega_\pm{}_{ij} \pm X_{m[i} \tilde k_{\pm j]} \mp \frac{1}{2} \hodge{Y}_{m[i|pq|}\pi_\pm{}^{pq}{}_{j]} \mp \frac{1}{2} W_m \omega_{\pm ij} \\
		&+ Z_{m[i|p|}\omega_\pm{}^p{}_{j]} + \frac{1}{2} P_{mp} \pi_\pm{}^p{}_{ij} \\
		=& \mp A^{-1} \partial_{m} A \, \tilde\omega_\pm{}_{ij} + \frac{1}{4!} \hodge{X}_{mijpqr} \pi_\pm{}^{pqr} \pm \frac{1}{2} \delta_{m[i} X_{j]p} \tilde k_\pm{}^p\pm \frac{3}{4} X_{[mi} \tilde k_{\pm j]}  \\
		&  \mp \frac{1}{12} \delta_{m[i} \hodge{Y}_{j]pqr} \pi_\pm{}^{pqr} \mp \frac{3}{8} \hodge{Y}_{[mi|pq|}\pi_\pm{}^{pq}{}_{j]} + \frac{1}{4} Y_{mijp} \tilde k_\pm{}^p \\
		&- \frac{1}{2} S \delta_{m[i} \tilde k_{\pm j]} \mp \frac{1}{4} N \pi_{\pm mij} - \frac{1}{4} \delta_{m[i}P_{|pq|} \pi_\pm{}^{pq}{}_{j]} + \frac{3}{4} P_{[m|p|}\pi_\pm{}^p{}_{ij]}~,
	\end{split}
\end{equation}

\begin{equation}
	\begin{split}
		{\cal D}^{\cal F}_m \pi_{\pm}{}_{ijk} \defeq& \nabla_m \pi_{\pm}{}_{ijk} \pm \frac{3}{2} X_{m[i}\omega_{\pm jk]} - \frac{3}{2} Y_{m[ij|p|}\omega_\pm{}^p{}_{k]}\mp \frac{3}{2} \hodge{Y}_{m[ij|p|} \tilde\omega_\pm{}^p{}_{k]} \\
		&\pm \frac{3}{4} \hodge{Z}_{m[ij|pq|} \pi_\pm{}^{pq}{}_{k]} - \frac{3}{2} Z_{m[ij} \tilde k_{\pm k]} - \frac{3}{2} P_{m[i} \tilde\omega_{\pm jk]}\\
		=& \mp A^{-1} \partial_{m} A \, \pi_{\pm}{}_{ijk} - \frac{1}{8} \hodge{X}_{mijkpq} \tilde \omega_\pm{}^{pq} \pm \frac{3}{2} \delta_{m[i}X_{j|p|}\omega_\pm{}^p{}_{k]} \pm \frac{3}{2} X_{[mi}\omega_{\pm jk]} \\
		&+ \frac{3}{8} \delta_{m[i}Y_{jk]pq}\omega_\pm{}^{pq} - Y_{[mij|p|}\omega_\pm{}^p{}_{k]} \pm \frac{3}{8} \delta_{m[i}\hodge{Y}_{jk]pq}\tilde\omega_\pm{}^{pq} \mp \hodge{Y}_{[mij|p|} \tilde\omega_\pm{}^p{}_{k]} \\
		&- \frac{3}{4} S \delta_{m[i}\omega_{\pm jk]} \pm\frac{1}{4} \delta_{m[i} \hodge{Z}_{jk]pqr} \pi_\pm{}^{pqr} \pm \hodge{Z}_{[mij|pq|}\pi_\pm{}^{pq}{}_{k]} \pm \frac{3}{4} N \delta_{m[i} \tilde \omega_{\pm jk]} \\
		&\pm \frac{1}{8} \hodge{P}_{mijkpq} \omega_\pm{}^{pq} - \frac{3}{2} \delta_{m[i}P_{j|p|} \tilde \omega_\pm{}^p{}_{k]} - \frac{3}{2} P_{[mi} \tilde\omega_{\pm jk]}~,
	\end{split}
\end{equation}
where for simplicity we have suppressed the $r,s$ indices on the form bilinears that label the different Killing spinors. Again the above conditions on the form bilinears have been organised as those of a TCFH in (\ref{tcfhcon}).

As it is apparent from the analysis above,  the domain of the minimal TCFH connection $\mathcal{D}^\mathcal{F}$ can be identified with $\Omega^*(M^8)$. This is the span of $\phi$ and the Hodge dual of $\tilde\phi$ form bilinears\footnote{Note that $\zeta$ and $\tilde\zeta$ are Hodge duals and so only $\zeta$ is chosen to belong in the basis.}. This domain factorises into the space of symmetric form bilinears, \eqref{AdS2_Sym_Bilinears} and the space of skew-symmetric form bilinears, \eqref{AdS2_Skew_Bilinears}. This can be understood as follows.  The spinors $\eta_\pm$ can be viewed as Majorana $\mathfrak{spin}(8)$ spinors. The product of two Majorana $\mathfrak{spin}(8)$ representations, $\Delta_{16}$,  decomposes as
\begin{equation}
	\otimes^2 \Delta_{16} = \Lambda^*(\mathbb{R}^8)~,
\end{equation}
and so the space of form bilinears spans all forms over $M^8$, where $\oplus_{k=0}^4 \Lambda^k(\mathbb{R}^8)$ is associated with the span of $\phi$ form bilinears while $\oplus_{k=5}^8 \Lambda^k(\mathbb{R}^8)$ is associated with the span of the Hodge duals of the $\tilde\phi$ form bilinears. Indeed, we note that $\textrm{dim}(\otimes^2 \Delta_{16}) = 2^4 \cdot 2^4 = \textrm{dim}(\Lambda^*(\mathbb{R}^8))$. Thus $\mathcal{D}^\mathcal{F}$ acts on the space of all forms on $M^8$. However, we see that the minimal TCFH connection preserves the subspaces of  form bilinears that are symmetric and skew-symmetric in the exchange of the two Killing spinors respectively, i.e. it preserves the symmetrised $S^2(\Delta_{16})$ and skew-symmetrised $\Lambda^2(\Delta_{16})$ subspaces of $\otimes^2 \Delta_{16}$. Therefore, the reduced holonomy of $\mathcal{D}^\mathcal{F}$ will be contained within the connected component\footnote{The reduced holonomy of a connection is by definition connected. So from now on when we refer to a group in the context of reduced holonomy we shall  consider only its connected component even if this is not explicitly mentioned. } of $GL(136) \times GL(120)$. However, the reduced holonomy of the minimal TCFH connection reduces further to $GL(134) \times GL(120)$ as it acts with partial derivatives on the scalars $f$ and $\tilde{f}$ and so their contribution to the holonomy is trivial.

\section{The TCFH of warped AdS\textsubscript{3} backgrounds}

\subsection{Fields and Killing spinors}

The bosonic fields of warped AdS$_3$ backgrounds, $AdS_3 \times_w M^7$, with internal space $M^7$  of  massive IIA supergravity  can be expressed as
\bea
		&&g = 2\, \e^+ \e^- + (\e^z)^2 + g(M^7)~,~~~
		G =  \e^+ \wedge \e^- \wedge \e^z  \wedge X + Y~, \cr
&&  H =  W\, \e^+ \wedge \e^- \wedge \e^z + Z~, ~~~
		F = F~, \quad S = m e^\Phi~, \quad \Phi = \Phi~,
\label{ads3back}
	\eea
where $m$ is a constant, $g(M^7)$ is a metric on $M^7$, $\Phi, W \in C^\infty(M^7)$, $X \in \Omega^1(M^7)$, $F \in \Omega^2(M^7)$, $Z \in \Omega^3(M^7)$ and $Y \in \Omega^4(M^7)$.  Note that the Bianchi identities imply that either $S = 0$ or $W = 0$.  Further,
\begin{equation}
	\begin{gathered}
		\e^+ = \dd u, \quad \e^- = \dd r -   \frac{2}{\ell} r  \dd z - 2 r A^{-1} \dd A, \quad \e^z=A \dd z~, \quad \e^i = e^i{}_J \dd y^J~,
	\end{gathered}
\end{equation}
is a  pseudo-orthonormal frame on $AdS_3 \times_w M^7$ with $g(M^7) = \delta_{ij} \e^i \e^j$, where $y$ are the coordinates of the internal space and  $(u, r, z)$ are the remaining coordinates of spacetime. After a coordinate transformation, the spacetime metric takes the standard warped form $g=A^2 g_\ell(AdS_3)+ g(M^7)$ with warp factor  $A$, $A\in C^\infty(M^7)$,  where $g_\ell(AdS_3)$ is the standard metric on AdS$_3$ of radius $\ell$.

As in the previous case, the KSEs of warped AdS\textsubscript{3} backgrounds can be integrated over the coordinates $(u,r,z)$, see \cite{bgpiia}.  The Killing spinors can be written schematically as $\epsilon=\epsilon(u,r,z, \sigma_\pm, \tau_\pm)$, where the spinors $\sigma_\pm$ and $\tau_\pm$ depend only on the coordinates of $M^7$ and satisfy $\Gamma_\pm\sigma_\pm=\Gamma_\pm \tau_\pm=0$.  Moreover, the gravitino KSE implies that $\mathcal{D}_m^{(\pm)} \chi_\pm = 0$, where
\begin{equation}
	\begin{split}
			\mathcal{D}_m^{(\pm)} = \nabla_m \, \pm & \frac{1}{2} A^{-1} \partial_{m} A \, + \frac{1}{8} \slashed{Z}_{m} \Gamma_{11}  + \frac{1}{8} S \Gamma_{m} \\&+ \frac{1}{16} \slashed{F}\Gamma_{m}\Gamma_{11}  + \frac{1}{192} \slashed{Y} \Gamma_{m}  \pm \frac{1}{8} \slashed{X} \Gamma_{z m}~,
	\end{split}
\label{supconeads3}
\end{equation}
is the supercovariant derivative along the internal space $M^7$, $m=1,\dots, 7$ ,   $\nabla$ is the spin connection associated with the metric $g(M^7)$ and $\chi_\pm$ stands for either $\sigma_\pm$ or $\tau_\pm$.

The Killing spinors $\chi_\pm$  satisfy two  algebraic KSEs \cite{bgpiia} in addition to the gravitino KSE along $M^7$. One of these is induced by the dilatino KSE of massive IIA supergravity.  The other arises during the integration of the gravitino KSE of massive IIA supergravity over the $z$ spacetime coordinate.  We shall not describe these here as they are not essential for the description of the TCFH on $M^7$.  However, they are necessary for the correct counting of Killing spinors in the examples that follow and a brief mention will be made where it is needed.

For warped AdS$_3$ backgrounds, the $\sigma_\pm$ and $\tau_\pm$ spinors are independent, i.e. there is no a priori Clifford algebra operation that relates the $\sigma_\pm$ solutions of the KSEs to the $\tau_\pm$ ones. A well known consequence of this is that the symmetry superalgebra of warped AdS$_3$ backgrounds factorises into a left and right sector that commute with each other. As we shall mention later, this is no longer the case for warped AdS$_k$, $k>3$, backgrounds where the $\sigma_\pm$ and $\tau_\pm$ Killing spinors are related with Clifford algebra operations.

\subsection{The TCFH on $M^7$}

Given Killing spinors $\chi_\pm^r$ and $\chi_\pm^s$, the form bilinears on $M^7$ can be constructed as for AdS$_2$ backgrounds in (\ref{fbi}) with $\eta_\pm$ replaced with $\chi_\pm$.  However there are differences. One is   that now $\e^i$ is an orthonormal frame on $M^7$ instead  on $M^8$ as was the case for AdS$_2$ backgrounds. The other is   that one can also insert in addition to $\Gamma_{11}$ the gamma matrix $\Gamma_z$ in the form bilinears.  Again, the reality condition on $\chi_\pm$ implies that the form bilinears are either symmetric  or skew-symmetric in the exchange of $\chi_\pm^r$ and $\chi_\pm^s$.

A basis in the space of  form bilinears\footnote{ The TCFHs  associated  with the form bilinears constructed from the pairs $(\sigma_+, \tau_+)$ and $(\sigma_+, \sigma_+)$ (and $(\sigma_-, \tau_-)$   and $(\sigma_-, \sigma_-)$) are identical  as the supercovariant connection (\ref{supconeads3}) on $\sigma_\pm$ is identical to that on $\tau_\pm$. So it is  sufficient to consider only the TCFHs of the form bilinears constructed from the pairs $(\sigma_+,\sigma_+)$ and $(\sigma_-,\sigma_-)$.}
on $M^7$, up to Hodge duality, which are symmetric in the exchange of Killing spinors $\chi_\pm^r$ and $\chi_\pm^s$ is
\begin{equation} \label{AdS3_Sym_Bilinears}
	\begin{gathered}
		f^{rs}_\pm = \H{\chi^r_\pm}{\chi^s_\pm} , \quad  \tilde{f}^{rs}_\pm = \H{\chi^r_\pm}{\Gamma_{11} \chi^s_\pm} , \quad	\hat f^{rs}_{\pm} = \H{\chi^r_\pm}{\Gamma_ z\chi^s_\pm} , \\
		k^{rs}_\pm = \H{\chi^r_\pm}{\Gamma_ i\chi^s_\pm} \, \e^i, \quad	 \mathring \omega^{rs}_{\pm} = \frac{1}{2} \H{\chi^r_\pm}{\Gamma_ {ijz}\Gamma_{11} \chi^s_\pm} \, \e^i \wedge \e^j , \\
		\tilde{\pi}^{rs}_\pm = \frac{1}{3!} \H{\chi^r_\pm}{\Gamma_ {ijk}\Gamma_{11} \chi^s_\pm} \, \e^i \wedge \e^j \wedge \e^k, \quad		\hat \pi^{rs}_{\pm} = \frac{1}{3!} \H{\chi^r_\pm}{\Gamma_{ijkz}\chi^s_\pm} \, \e^{i} \wedge \e^j \wedge \e^k , \\
		\mathring \pi^{rs}_{\pm} = \frac{1}{3!} \H{\chi^r_\pm}{\Gamma_{ijkz}\Gamma_{11}\chi^s_\pm} \, \e^{i} \wedge \e^j \wedge \e^k~.
	\end{gathered}
\end{equation}
The computation of the TCFH follows the steps described in section \ref{sec:TCFH}. In particular the TCFH expressed in terms of the minimal  connection, $\mathcal{D}^\mathcal{F}$, is

\begin{equation}
    \begin{split}
	    {\cal D}^{\cal F}_m f_\pm \defeq& \nabla_m f_\pm \\
        =& \mp A^{-1} \partial_{m} A \, f_\pm  - \frac{1}{4}Sk_\pm{}_m -\frac{1}{8}F_{pq}\tilde{\pi}_\pm{}^{pq}{}_m \pm \frac{1}{8}\hodge{Y}_{mpq}\mathring \omega_{\pm}{}^{pq} \pm \frac{1}{4}X_m \hat f_{\pm}~,
    \end{split}
\end{equation}

\begin{equation}
    \begin{split}
	    {\cal D}^{\cal F}_m \tilde{f}_\pm \defeq& \nabla_m \tilde{f}_\pm \\
        =& \mp A^{-1} \partial_{m} A \, \tilde{f}_\pm - \frac{1}{4} F_{mp}k_\pm{}^p - \frac{1}{4!} Y_{mpqr} \tilde{\pi}_\pm{}^{pqr} \mp \frac{1}{4} X_p \mathring \omega_{\pm}{}^p{}_m~,
    \end{split}
\end{equation}

\begin{equation}
    \begin{split}
	    {\cal D}^{\cal F}_m \hat f_{\pm} \defeq& \nabla_m \hat f_{\pm} \\
        =& \mp A^{-1} \partial_{m} A \, \hat f_{\pm} -\frac{1}{4} Z_{mpq}\mathring \omega_{\pm}{}^{pq} + \frac{1}{8} F_{pq}\mathring \pi_{\pm}{}^{pq}{}_m - \frac{1}{4!}Y_{mpqr}\hat \pi_{\pm}{}^{pqr} \pm \frac{1}{4}X_m f_\pm~,
    \end{split}
\end{equation}

\begin{equation}
	\begin{split}
		{\cal D}^{\cal F}_m k_\pm{}_i \defeq& \nabla_m k_\pm{}_i + \frac{1}{4}Z_{mpq}\tilde{\pi}_\pm{}^{pq}{}_i \mp \frac{1}{4}\hodge{Y}_{mpq}\mathring \pi_{\pm}{}^{pq}{}_i\\
		=& \mp A^{-1} \partial_{m} A \, k_\pm{}_i \, -\frac{1}{4}\delta_{mi} Sf_\pm \mp \frac{1}{4!} \hodge{F}_{mipqr}\hat \pi_{\pm}{}^{pqr} + \frac{1}{4} F_{mi}\tilde{f}_\pm  \\
		& \mp \frac{1}{4!} \delta_{mi} \hodge{Y}_{pqr} \mathring \pi_{\pm}{}^{pqr} \mp \frac{1}{4} \hodge{Y}_{[m|pq|}\mathring \pi_{\pm}{}^{pq}{}_{i]} \pm \frac{1}{4}X_{p}\hat \pi_{\pm}{}^{p}{}_{mi}~,
	\end{split}
\end{equation}

\begin{equation}
	\begin{split}
		{\cal D}^{\cal F}_m \mathring \omega_{\pm}{}_{ij} \defeq& \nabla_m \mathring \omega_{\pm}{}_{ij} \mp \frac{1}{2}\hodge{Z}_{m[i|pq|}\tilde\pi_\pm{}^{pq}{}_{j]} - \frac{1}{2} F_{mp}\hat \pi_{\pm}{}^p{}_{ij} + \frac{1}{2} Y_{m[i|pq|} \mathring \pi_{\pm}{}^{pq}{}_{j]} \\
		=& \mp A^{-1} \partial_{m} A \, \mathring \omega_{\pm}{}_{ij} \mp \frac{1}{6} \delta_{m[i}\hodge{Z}_{j]pqr}\tilde\pi_\pm{}^{pqr} \mp \frac{3}{4}\hodge{Z}_{[mi|pq|}\tilde\pi_\pm{}^{pq}{}_{j]} \\
		 &+ \frac{1}{2} Z_{mij}\hat f_{\pm} -\frac{1}{4} S\mathring \pi_{\pm}{}_{mij} + \frac{1}{4}\delta_{m[i|}F_{pq|}\hat \pi_{\pm}{}^{pq}{}_{j]} \\
		 &- \frac{3}{4}F_{[m|p|}\hat \pi_{\pm}{}^p{}_{ij]} + \frac{1}{12} \delta_{m[i}Y_{j]pqr} \mathring \pi_{\pm}{}^{pqr} + \frac{3}{8} Y_{[mi|pq|}\mathring \pi_{\pm}{}^{pq}{}_{j]} \\
		&\pm \frac{1}{4} \hodge{Y}_{mij}f_\pm + \frac{1}{4!} \hodge{X}_{mijpqr}\hat \pi_{\pm}{}^{pqr} \mp \frac{1}{2} \delta_{m[i}X_{j]}\tilde{f}_\pm~,
	\end{split}
\end{equation}

\begin{equation}
	\begin{split}
		{\cal D}^{\cal F}_m \tilde{\pi}_{\pm}{}_{ijk} \defeq& \nabla_m \tilde{\pi}_{\pm}{}_{ijk} \mp \frac{3}{2} \hodge{Z}_{m[ij|p|}\mathring \omega_{\pm}{}^p{}_{k]} - \frac{3}{2}Z_{m[ij}k_\pm{}_{k]} \pm \frac{3}{4} \hodge{F}_{m[ij|pq|}\mathring \pi_{\pm}{}^{pq}{}_{k]} \\
		&\pm \frac{3}{2}\hodge{Y}_{m[i|p|}\hat \pi_{\pm}{}^p{}_{jk]} \pm \frac{1}{2}X_m \mathring \pi_{\pm}{}_{ijk} \\
		=& \mp A^{-1} \partial_{m} A \, \tilde{\pi}_{\pm}{}_{ijk} \mp 2 \hodge{Z}_{[mij|p|}\mathring \omega_{\pm}{}^p{}_{k]} \pm \frac{3}{4} \delta_{m[i}\hodge{Z}_{jk]pq}\mathring \omega_{\pm}{}^{pq}  \\
		&\pm \frac{1}{4!} \hodge{S}_{mijkpqr}\hat \pi_{\pm}{}^{pqr} \pm \frac{1}{8} \delta_{m[i}\hodge{F}_{jk]pqr}\mathring \pi_{\pm}{}^{pqr} \pm \frac{1}{2} \hodge{F}_{[mij|pq|}\mathring \pi_{\pm}{}^{pq}{}_{k]} \\
		&- \frac{3}{4}\delta_{m[i}F_{jk]}f_\pm \mp \frac{3}{4}\delta_{m[i}\hodge{Y}_{j|pq|}\hat \pi_{\pm}{}^{pq}{}_{k]} \pm \frac{3}{2} \hodge{Y}_{[mi|p|}\hat \pi_{\pm}{}^p{}_{jk]} \\
		&+ \frac{1}{4}Y_{mijk}\tilde{f}_\pm \pm X_{[m}\mathring \pi_{\pm}{}_{ijk]} \pm \frac{3}{4} \delta_{m[i|}X_{p|}\mathring \pi_{\pm}{}^p{}_{jk]}~,
	\end{split}
\end{equation}

\begin{equation}
	\begin{split}
		{\cal D}^{\cal F}_m \hat \pi_{\pm}{}_{ijk} \defeq& \nabla_m \hat \pi_{\pm}{}_{ijk} + \frac{3}{2} Z_{m[i|p|}\mathring \pi_{\pm}{}^p{}_{jk]} + \frac{3}{2}F_{m[i}\mathring \omega_{\pm}{}_{jk]} \mp \frac{3}{2}\hodge{Y}_{m[i|p|}\tilde\pi_\pm{}^p{}_{jk]} \\
		=& \mp A^{-1} \partial_{m} A \, \hat \pi_{\pm}{}_{ijk} \mp \frac{1}{4!} \hodge{S}_{mijkpqr}\tilde\pi_\pm{}^{pqr} \pm \frac{1}{4} \hodge{F}_{mijkp}k_\pm{}^p \\
		&+\frac{3}{2}F_{[mi}\mathring \omega_{\pm}{}_{jk]} + \frac{3}{2}\delta_{m[i}F_{j|p|}\mathring \omega_{\pm}{}^p{}_{k]} \pm \frac{3}{4} \delta_{m[i} \hodge{Y}_{j|pq|} \tilde\pi_\pm{}^{pq}{}_{k]} \mp \frac{3}{2} \hodge{Y}_{[mi|p|} \tilde\pi_\pm{}^p{}_{jk]} \\
		&+ \frac{1}{4}Y_{mijk}\hat f_{\pm}-\frac{1}{8}\hodge{X}_{mijkpq}\mathring \omega_{\pm}{}^{pq} \pm \frac{3}{2}\delta_{m[i}X_jk_{\pm}{}_{k]}~,
	\end{split}
\end{equation}

\begin{equation}
	\begin{split}
		{\cal D}^{\cal F}_m \mathring \pi_{\pm}{}_{ijk} \defeq& \nabla_m \mathring \pi_{\pm}{}_{ijk} + \frac{3}{2} Z_{m[i|p|}\hat \pi_{\pm}{}^p{}_{jk]} \pm \frac{3}{4} \hodge{F}_{m[ij|pq|}\tilde\pi_\pm{}^{pq}{}_{k]} - \frac{3}{2} Y_{m[ij|p|} \mathring \omega_{\pm}{}^p{}_{k]} \\
		&\mp \frac{3}{2} \hodge{Y}_{m[ij}k_{\pm}{}_{k]} \pm \frac{1}{2} X_m \tilde\pi_\pm{}_{ijk} \\
		=& \mp A^{-1} \partial_{m} A \, \mathring \pi_{\pm}{}_{ijk} - \frac{3}{4} S \delta_{m[i}\mathring \omega_{\pm}{}_{jk]} + \frac{3}{4} \delta_{m[i} F_{jk]}\hat f_{\pm} \pm \frac{1}{8} \delta_{m[i}\hodge{F}_{jk]pqr}\tilde\pi_\pm{}^{pqr} \\
		& \pm \frac{1}{2} \hodge{F}_{[mij|pq|} \tilde\pi_\pm{}^{pq}{}_{k]} + \frac{3}{8} \delta_{m[i}Y_{jk]pq}\mathring \omega_{\pm}{}^{pq} - Y_{[mij|p|} \mathring \omega_{\pm}{}^p{}_{k]} \mp \frac{3}{4} \delta_{m[i}\hodge{Y}_{jk]p}k_\pm{}^p \\
		&\mp \hodge{Y}_{[mij}k_\pm{}_{k]} \pm X_{[m}\tilde\pi_\pm{}_{ijk]} \pm \frac{3}{4} \delta_{m[i|} X_{p|}\tilde\pi_\pm{}^p{}_{jk]}~,
	\end{split}
\end{equation}
where for simplicity we have suppressed the $r,s$ indices on the form bilinears that label the different Killing spinors.

Similarly a basis in the space of Killing spinor bilinears of $AdS_3 \times_w M^7$, up to Hodge duality, which are skew-symmetric in the exchange of Killing spinors is
\begin{equation} \label{AdS3_Skew_Bilinears}
	\begin{gathered}
		\mathring f^{rs}_{\pm} = \H{\chi^r_\pm}{\Gamma_{z} \Gamma_{11} \chi^s_\pm}~, \quad \tilde k^{rs}_{\pm} = \H{\chi^r_\pm}{\Gamma_{i}\Gamma_{11}\chi^s_\pm} \, \e^i , \quad \hat k^{rs}_{\pm} = \H{\chi^r_\pm}{\Gamma_{iz}\chi^s_\pm} \, \e^i~, \\
		\mathring k^{rs}_{\pm} = \H{\chi^r_\pm}{\Gamma_{iz}\Gamma_{11}\chi^s_\pm} \, \e^i~, \quad \omega^{rs}_\pm = \frac{1}{2} \H{\chi^r_\pm}{\Gamma_{ij}\chi^s_\pm} \, \e^i \wedge \e^j~, \\
		\tilde{\omega}^{rs}_\pm = \frac{1}{2} \H{\chi^r_\pm}{\Gamma_{ij}\Gamma_{11}\chi^s_\pm} \, \e^i \wedge \e^j~, \quad \hat \omega^{rs}_{\pm} = \frac{1}{2} \H{\chi^r_\pm}{\Gamma_ {ijz} \chi^s_\pm} \, \e^i \wedge \e^j~, \\
		\pi^{rs}_\pm = \frac{1}{3!} \H{\chi^r_\pm}{\Gamma_ {ijk} \chi^s_\pm} \, \e^i \wedge \e^j \wedge \e^k~.
	\end{gathered}
\end{equation}
The associated TCFH on $M^7$ with respect to the minimal  connection, ${\cal D}^{\cal F}$, reads

\begin{equation}
	\begin{split}
		{\cal D}^{\cal F}_m \mathring f_{\pm} \defeq& \nabla_m \mathring f_{\pm} \\
		=& \mp A^{-1} \partial_{m} A \, \mathring f_{\pm} - \frac{1}{4} Z_{mpq} \hat \omega_{\pm}{}^{pq} - \frac{1}{4} S \mathring k_{\pm m} \\
		&+ \frac{1}{4} F_{mp} \hat k_{\pm}{}^p \mp \frac{1}{8} \hodge{Y}_{mpq} \omega_\pm{}^{pq} \mp \frac{1}{4} X_p \tilde\omega_{\pm}{}^p{}_m~,
	\end{split}
\end{equation}

\begin{equation}
	\begin{split}
		{\cal D}^{\cal F}_m \tilde k_\pm{}_i \defeq& \nabla_m \tilde k_\pm{}_i - \frac{1}{2} F_{mp} \omega_\pm{}^p{}_i \pm \frac{1}{2} X_m \mathring k_{\pm i} \\
		=& \mp A^{-1} \partial_{m} A \, \tilde k_\pm{}_i \, -\frac{1}{4} Z_{mpq} \pi_\pm{}^{pq}{}_i - \frac{1}{4} S \tilde\omega_{\pm mi} + \frac{1}{8} \delta_{mi} F_{pq} \omega_{\pm}{}^{pq} \\
		&- \frac{1}{2} F_{[m|p|} \omega_\pm{}^p{}_{i]} \mp \frac{1}{4} \hodge{Y}_{mip} \hat k_{\pm}{}^p - \frac{1}{8} Y_{mipq} \tilde\omega_\pm{}^{pq} \pm \frac{1}{4} \delta_{mi} X_p \mathring k_{\pm}{}^p \\
		 &\pm \frac{1}{2} X_{[m} \mathring k_{\pm i]}~,
	\end{split}
\end{equation}

\begin{equation}
	\begin{split}
		{\cal D}^{\cal F}_m \hat k_{\pm}{}_i \defeq& \nabla_m \hat k_{\pm}{}_i \\
		=& \mp A^{-1} \partial_{m} A \, \hat k_{\pm}{}_i \, - \frac{1}{2} Z_{mip} \mathring k_{\pm}{}^p - \frac{1}{4} S \hat \omega_{\pm mi} \mp \frac{1}{4!} \hodge{F}_{mipqr} \pi_{\pm}{}^{pqr} - \frac{1}{4} F_{mi} \mathring f_{\pm} \\
		&\pm \frac{1}{4} \hodge{Y}_{mip} \tilde k_\pm{}^p - \frac{1}{8} Y_{mipq} \hat \omega_{\pm}{}^{pq} \pm \frac{1}{4} X_p \pi_\pm{}^p{}_{mi}~,
	\end{split}
\end{equation}

\begin{equation}
	\begin{split}
		{\cal D}^{\cal F}_m \mathring k_{\pm}{}_i \defeq& \nabla_m \mathring k_{\pm}{}_i + \frac{1}{2} F_{mp} \hat \omega_{\pm}{}^p{}_i \pm \frac{1}{4} \hodge{Y}_{mpq} \pi_\pm{}^{pq}{}_i \pm \frac{1}{2} X_m \tilde k_{\pm i} \\
		=& \mp A^{-1} \partial_{m} A \, \mathring k_{\pm}{}_i \, - \frac{1}{2} Z_{mip} \hat k_{\pm}{}^p - \frac{1}{4} S \delta_{mi} \mathring f_{\pm} - \frac{1}{8} \delta_{mi} F_{pq} \hat \omega_{\pm}{}^{pq} \\
		&+ \frac{1}{2} F_{[m|p|} \hat \omega_{\pm}{}^p{}_{i]} \pm \frac{1}{4!} \delta_{mi} \hodge{Y}_{pqr} \pi_\pm{}^{pqr} \pm \frac{1}{4} \hodge{Y}_{[m|pq|} \pi_\pm{}^{pq}{}_{i]} \\
		&\pm \frac{1}{4} \delta_{mi} X_p \tilde k_{\pm}{}^p \pm \frac{1}{2} X_{[m} \tilde k_{\pm i]}~,
	\end{split}
\end{equation}

\begin{equation}
	\begin{split}
		{\cal D}^{\cal F}_m \omega_{\pm ij} \defeq& \nabla_m \omega_{\pm ij} + Z_{m[i|p|} \tilde \omega_\pm{}^p{}_{j]} + F_{m[i} \tilde k_{\pm j]} + \frac{1}{2} Y_{m[i|pq|} \pi_\pm{}^{pq}{}_{j]} \mp \frac{1}{2} X_m \hat \omega_{\pm ij} \\
		=& \mp A^{-1} \partial_{m} A \, \omega_{\pm ij} \, - \frac{1}{4} S \pi_{\pm mij} \pm \frac{1}{8} \hodge{F}_{mijpq} \hat \omega_{\pm}{}^{pq} \\
		&+ \frac{1}{2} \delta_{m[i} F_{j]p} \tilde k_\pm{}^p + \frac{3}{4} F_{[mi} \tilde k_{\pm j]} + \frac{1}{12} \delta_{m[i} Y_{j]pqr} \pi_\pm{}^{pqr} \\
		&+ \frac{3}{8} Y_{[mi|pq|} \pi_\pm{}^{pq}{}_{j]} \mp \frac{1}{4} \hodge{Y}_{mij} \mathring f_{\pm} \mp \frac{1}{2} \delta_{m[i} X_{|p|} \hat \omega_{\pm}{}^p{}_{j]} \\
		&\mp \frac{3}{4} X_{[m} \hat \omega_{\pm ij]}~,
	\end{split}
\end{equation}

\begin{equation}
	\begin{split}
		{\cal D}^{\cal F}_m \tilde\omega_{\pm ij} \defeq& \nabla_m \tilde\omega_{\pm ij} + Z_{m[i|p|} \omega_\pm{}^p{}_{j]} + \frac{1}{2} F_{mp} \pi_\pm{}^p{}_{ij} \pm \hodge{Y}_{m[i|p|} \hat \omega_{\pm}{}^p{}_{j]} \\
		=& \mp A^{-1} \partial_{m} A \, \tilde\omega_{\pm ij} \, - \frac{1}{2} S \delta_{m[i} \tilde k_{\pm j]} - \frac{1}{4} \delta_{m[i} F_{|pq|} \pi_\pm{}^{pq}{}_{j]} \\
		&+ \frac{3}{4} F_{[m|p|} \pi_\pm{}^p{}_{ij]} \mp \frac{1}{4} \delta_{m[i} \hodge{Y}_{j]pq} \hat \omega_{\pm}{}^{pq} \pm \frac{3}{4} \hodge{Y}_{[mi|p|} \hat \omega_{\pm}{}^p{}_{j]} + \frac{1}{4} Y_{mijp} \tilde k_{\pm}{}^p \\
		&- \frac{1}{4!} \hodge{X}_{mijpqr} \pi_\pm{}^{pqr} \mp \frac{1}{2} \delta_{m[i} X_{j]} \mathring f_{\pm}~,
	\end{split}
\end{equation}

\begin{equation}
	\begin{split}
		{\cal D}^{\cal F}_m \hat \omega_{\pm ij} \defeq& \nabla_m \hat \omega_{\pm ij} \mp \frac{1}{2} \hodge{Z}_{m[i|pq|} \pi_\pm{}^{pq}{}_{j]} - F_{m[i} \mathring k_{\pm j]} \mp \hodge{Y}_{m[i|p|} \tilde \omega_\pm{}^p{}_{j]} \mp \frac{1}{2} X_m \omega_{\pm ij} \\
		=& \mp A^{-1} \partial_{m} A \, \hat \omega_{\pm ij} \, \mp	 \frac{1}{6} \delta_{m[i} \hodge{Z}_{j]pqr} \pi_\pm{}^{pqr} \mp \frac{3}{4} \hodge{Z}_{[mi|pq|} \pi_\pm{}^{pq}{}_{j]} \\
		&+ \frac{1}{2} Z_{mij} \mathring f_{\pm} - \frac{1}{2} S \delta_{m[i} \hat k_{\pm j]} \pm \frac{1}{8} \hodge{F}_{mijpq} \omega_\pm{}^{pq} - \frac{1}{2} \delta_{m[i} F_{j]p} \mathring k_{\pm}{}^p \\
		&- \frac{3}{4} F_{[mi} \mathring k_{\pm j]} + \frac{1}{4} Y_{mijp} \hat k_{\pm}{}^p \pm \frac{1}{4} \delta_{m[i} \hodge{Y}_{j]pq} \tilde \omega_\pm{}^{pq} \mp \frac{3}{4} \hodge{Y}_{[mi|p|} \tilde\omega_\pm{}^p{}_{j]} \\
		&\mp \frac{1}{2} \delta_{m[i} X_{|p|} \omega_\pm{}^p{}_{j]} \mp \frac{3}{4} X_{[m} \omega_{\pm ij]}~,
	\end{split}
\end{equation}

\begin{equation}
	\begin{split}
		{\cal D}^{\cal F}_m \pi_{\pm ijk} \defeq& \nabla_m \pi_{\pm ijk} \mp \frac{3}{2} \hodge{Z}_{m[ij|p|} \hat \omega_{\pm}{}^p{}_{k]} - \frac{3}{2} Z_{m[ij} \tilde k_{\pm k]} - \frac{3}{2} F_{m[i} \tilde \omega_{\pm jk]} \\
		&- \frac{3}{2} Y_{m[ij|p|} \omega_\pm{}^p{}_{k]} \pm \frac{3}{2} \hodge{Y}_{m[ij} \mathring k_{\pm k]} \\
		=& \mp A^{-1} \partial_{m} A \, \pi_{\pm ijk} \, \pm \frac{3}{4} \delta_{m[i} \hodge{Z}_{jk]pq} \hat \omega_{\pm}{}^{pq} \mp 2 \hodge{Z}_{[mij|p|} \hat \omega_{\pm}{}^p{}_{k]} \\
		&- \frac{3}{4} S \delta_{m[i} \omega_{\pm jk]} \pm \frac{1}{4} \hodge{F}_{mijkp} \hat k_{\pm}{}^p - \frac{3}{2} \delta_{m[i} F_{j|p|} \tilde \omega_{\pm}{}^p{}_{k]} - \frac{3}{2} F_{[mi} \tilde \omega_{\pm jk]} \\
		&+ \frac{3}{8} \delta_{m[i} Y_{jk]pq} \omega_\pm{}^{pq} - Y_{[mij|p|} \omega_\pm{}^p{}_{k]} \pm \frac{3}{4} \delta_{m[i} \hodge{Y}_{jk]p} \mathring k_{\pm}{}^p \pm \hodge{Y}_{[mij} \mathring k_{\pm k]} \\
		&+ \frac{1}{8} \hodge{X}_{mijkpq} \tilde \omega_\pm{}^{pq} \pm \frac{3}{2} \delta_{m[i} X_{j} \hat k_{\pm k]}~,
	\end{split}
\end{equation}
where, again, for simplicity we have suppressed the $r,s$ indices on the form bilinears\footnote{From now on, we shall always suppress the $r,s$ indices on the form bilinears that label the different Killing spinors in all the TCFHs below.}.

Upon using Hodge duality on $M^7$, the domain of $\mathcal{D}^\mathcal{F}$ can be identified with  $\Omega^*(M^7)\oplus \Omega^*(M^7)$. Moreover it is clear from the TCFH above that the domain of  $\mathcal{D}^\mathcal{F}$ factorises into the space of symmetric form bilinears, \eqref{AdS3_Sym_Bilinears}, and the space of skew-symmetric form bilinears, \eqref{AdS3_Skew_Bilinears}.  To understand this observe  that the 16-dimensional Majorana representation, $\Delta_{16}$, of $\mathfrak{spin}(8)$ decomposes under $\mathfrak{spin}(7)$ into a sum of two 8-dimensional Majorana representations, $\Delta_{8}$. In turn  the product of two $\Delta_{16}$  viewed as representations of $\mathfrak{spin}(7)$ decompose as
\begin{equation}
	\otimes^2 \Delta_{16} = \Lambda^*(\mathbb{R}^7) \oplus \Lambda^*(\mathbb{R}^7)~.
\end{equation}
 Indeed, we note that $\textrm{dim}(\otimes^2 \Delta_{16}) = 2^4 \cdot 2^4 = 2\, \textrm{dim}(\Lambda^*(\mathbb{R}^7))$. However, we see that the minimal TCFH connection preserves the symmetrised $S^2(\Delta_{16})$ and skew-symmetrised $\Lambda^2(\Delta_{16})$ subspaces of $\otimes^2 \Delta_{16}$. Therefore, the reduced holonomy of $\mathcal{D}^\mathcal{F}$ will be contained within $GL(136) \times GL(120)$. However, the reduced holonomy of the minimal TCFH connection reduces further to a subgroup of $GL(133) \times SO(7) \times GL(112)$ as it acts with  partial derivatives on the scalars $f$, $\tilde{f}$, $\hat f$ and $\mathring{f}$, and with the Levi-Civita  connection on $\hat{k}$.

\section{The TCFH of warped AdS\textsubscript{4} backgrounds}

\subsection{Fields and Killing spinors}

As in the previous cases, the bosonic fields  of warped AdS$_4$  backgrounds, AdS$_4 \times_w M^6$, with internal space $M^6$ of massive IIA supergravity can be expressed as
\bea
		&&g = 2\, \e^+ \e^- + (\e^z)^2+ (\e^x)^2  + g(M^6)~, \quad
		G =  X \, \e^+ \wedge \e^- \wedge \e^z \wedge \e^x + Y~, \cr
&&
		H = H~, \quad F = F~, \quad S = m e^\Phi~, \quad \Phi = \Phi~,
	\eea
where $g(M^6)$ is a metric on $M^6$, $m$ is a constant, $\Phi, X \in C^\infty(M^6)$, $F \in \Omega^2(M^6)$, $H \in \Omega^3(M^6)$ and $Y \in \Omega^4(M^6)$. Further,
\begin{equation}
	\begin{gathered}
		\e^+ = \dd u, \quad \e^- = \dd r - r \frac{2}{\ell} \dd z - 2 r A^{-1} \dd A~, \quad \e^z= A  \dd z~, \quad \e^x= A e^{z/\ell} \dd x~, \quad \e^i = e^i{}_J \dd y^J~,
	\end{gathered}
\end{equation}
is a pseudo-orthonormal frame on AdS$_4 \times_w M^6$ with $g(M^6)=\delta_{ij}\e^i \e^j$, where $y$ are the coordinates of $M^6$ and  $(u, r, z, x)$ are the remaining coordinates of spacetime. As in previous cases after a coordinate transformation the spacetime metric $g$ can be put into standard warped form $g=A^2 g_\ell(AdS_4)+g(M^6)$, where $A$ is the warp factor, $A\in C^\infty(M^6)$, and  $g_\ell(AdS_4)$ is the standard metric on AdS$_4$ with radius $\ell$.

Integrating the KSEs of massive IIA supergravity along the coordinates $(u,r,z,x)$, one finds that the Killing spinors can be expressed as $\epsilon=\epsilon(u,r,z,x, \sigma_\pm, \tau_\pm)$, where $\sigma_\pm$ and $\tau_\pm$ are spinors that depend only on the coordinates of $M^6$ and $\Gamma_\pm\sigma_\pm=\Gamma_\pm\tau_\pm=0$ \cite{bgpiia}. Furthermore, the gravitino KSE restricts $\sigma_\pm$ and $\tau_\pm$ along $M^6$ as $\mathcal{D}_m^{(\pm)} \chi_\pm = 0$, where $\chi_\pm$ stands for either $\sigma_\pm$ or $\tau_\pm$ and
\begin{equation}
	\begin{split}
			\mathcal{D}_m^{(\pm)} = \nabla_m \, \pm & \frac{1}{2} A^{-1} \partial_{m} A  +\frac{1}{8} \slashed{H}_{m}\Gamma_{11} + \frac{1}{8} S \Gamma_{m} \\
			+& \frac{1}{16} \slashed{F} \Gamma_{m} \Gamma_{11} + \frac{1}{192} \slashed{Y} \Gamma_{m} \mp \frac{1}{8} X \Gamma_{z x m}~,
	\end{split}
\end{equation}
with $\nabla_m$, $m=1,\dots6$, the spin connection of $g(M^6)$. The Killing spinors satisfy two additional algebraic KSEs. One is associated to the dilatino KSE of massive IIA supergravity and the other arises as a consequence of the integration of the gravitino KSE over $z$. Both are essential for identifying  the Killing spinors of a AdS$_4$ background but they do not contribute in the computation of TCFH on $M^6$. As a result will not be summarised here.

Unlike for warped AdS$_3$ backgrounds, the $\sigma_\pm$ and $\tau_\pm$ Killing spinors are related by a Clifford algebra operation. In particular, if $\sigma_\pm$ is a Killing spinor, then $\Gamma_{zx}\sigma_\pm$ is a $\tau_\pm$ Killing spinor, i.e. it solves all three Killing spinor equations that the $\tau_\pm$ Killing spinors satisfy  \cite{bgpiia}. Using this, one can demonstrate that the Killing spinors of AdS$_4$ backgrounds come in multiples of four.

\subsection{The TCFH of $M^6$}

The computation of the TCFH of warped AdS$_4$ backgrounds is similar to that of warped AdS$_2$ and AdS$_3$ cases that have already been described in some detail. Because of this we shall be brief. A basis in the space of Killing spinor form bilinears\footnote{We could have considered a more general class of bilinears like for example those that contain either a single insertion of $\Gamma_z$ or a single insertion of $\Gamma_x$, i.e. $\H{\chi^r_\pm}{\Gamma_z\chi^s_\pm}$ and $\H{\chi^r_\pm}{\Gamma_x\chi^s_\pm}$ for scalars and similarly for higher degree forms.  However, the choices of form bilinears below will suffice.} on $ M^6$, up to Hodge duality, which are symmetric in the exchange of Killing spinors $\chi^r_\pm$ and $\chi_\pm^s$ is
\begin{equation} \label{AdS4_Sym_Bilinears}
	\begin{gathered}
		f^{rs}_\pm = \H{\chi^r_\pm}{\chi^s_\pm}~, \quad  \tilde{f}^{rs}_\pm = \H{\chi^r_\pm}{\Gamma_{11} \chi^s_\pm}~, \quad	k^{rs}_\pm = \H{\chi^r_\pm}{\Gamma_ i\chi^s_\pm} \, \e^i~,  \\
		\mathring k^{rs}_{\pm} = \H{\chi^r_\pm}{\Gamma_ {izx}\Gamma_{11} \chi^s_\pm} \, \e^i~, \quad \hat \omega^{rs}_{\pm} = \frac{1}{2} \H{\chi^r_\pm}{\Gamma_{ijzx}\chi^s_\pm} \, \e^{i} \wedge \e^j~, \\
		\mathring \omega^{rs}_{\pm} = \frac{1}{2} \H{\chi^r_\pm}{\Gamma_{ijzx}\Gamma_{11}\chi^s_\pm} \, \e^{i} \wedge \e^j~, \quad \tilde{\pi}^{rs}_\pm = \frac{1}{3!} \H{\chi^r_\pm}{\Gamma_ {ijk}\Gamma_{11} \chi^s_\pm} \, \e^i \wedge \e^j \wedge \e^k~,
	\end{gathered}
\end{equation}
where again $\chi_\pm$ stands for either $\sigma_\pm$ or $\tau_\pm$.
After some computation, the TCFH   is
\begin{equation}
	\begin{split}
		{\cal D}^{\cal F}_m f_\pm \defeq& \nabla_m f_\pm \\
		=& \mp A^{-1} \partial_{m} A \, f_\pm  - \frac{1}{4}Sk_\pm{}_m -\frac{1}{8}F_{pq}\tilde{\pi}_\pm{}^{pq}{}_m \mp \frac{1}{4}\hodge{Y}_{mp}\mathring k_{\pm}{}^{p}~,
	\end{split}
\end{equation}

\begin{equation}
	\begin{split}
		{\cal D}^{\cal F}_m \tilde{f}_\pm \defeq& \nabla_m \tilde{f}_\pm \\
		=& \mp A^{-1} \partial_{m} A \, \tilde{f}_\pm - \frac{1}{4} F_{mp}k_\pm{}^p - \frac{1}{4!} Y_{mpqr} \tilde{\pi}_\pm{}^{pqr} \mp \frac{1}{4} X\mathring k_{\pm}{}_m~,
	\end{split}
\end{equation}

\begin{equation}
	\begin{split}
		{\cal D}^{\cal F}_m k_\pm{}_i \defeq& \nabla_m k_\pm{}_i + \frac{1}{4}H_{mpq}\tilde{\pi}_\pm{}^{pq}{}_i \mp \frac{1}{2} \hodge{Y}_{mp}\mathring \omega_{\pm}{}^p{}_i	\\
		=& \mp A^{-1} \partial_{m} A \, k_\pm{}_i \, -\frac{1}{4}\delta_{mi} Sf_\pm \pm \frac{1}{8} \hodge{F}_{mipq}\hat \omega_{\pm}{}^{pq} + \frac{1}{4} F_{mi}\tilde{f}_\pm \\
		&\pm \frac{1}{8} \delta_{mi}\hodge{Y}_{pq}\mathring \omega_{\pm}{}^{pq}  \mp \frac{1}{2}\hodge{Y}_{[m|p|}\mathring \omega_{\pm}{}^p{}_{i]} \mp \frac{1}{4}X\hat \omega_{\pm}{}_{mi}~,
	\end{split}
\end{equation}

\begin{equation}
	\begin{split}
		{\cal D}^{\cal F}_m \mathring k_{\pm}{}_i \defeq& \nabla_m \mathring k_{\pm}{}_i \mp \frac{1}{4} \hodge{H}_{mpq} \tilde\pi_\pm{}^{pq}{}_i - \frac{1}{2} F_{mp}\hat \omega_{\pm}{}^p{}_i \\
		=& \mp A^{-1} \partial_{m} A \, \mathring k_{\pm}{}_i \, \mp \frac{1}{12} \delta_{mi} \hodge{H}_{pqr} \tilde\pi_\pm{}^{pqr} \mp \frac{1}{2} \hodge{H}_{[m|pq|}\tilde\pi_\pm{}^{pq}{}_{i]} \\
		&-\frac{1}{4}S\mathring \omega_{\pm}{}_{mi} + \frac{1}{8}\delta_{mi} F_{pq} \hat \omega_{\pm}{}^{pq} - \frac{1}{2}F_{[m|p|}\hat \omega_{\pm}{}^p{}_{i]} \\
		&\mp \frac{1}{4}\hodge{Y}_{mi}f_\pm - \frac{1}{8}Y_{mipq} \mathring \omega_{\pm}{}^{pq} \pm \frac{1}{4}X\delta_{mi}\tilde{f}_\pm~,
	\end{split}
\end{equation}

\begin{equation}
	\begin{split}
		{\cal D}^{\cal F}_m \hat \omega_{\pm}{}_{ij} \defeq& \nabla_m \hat \omega_{\pm}{}_{ij} + H_{m[i|p|}\mathring \omega_{\pm}{}^p{}_{j]} + F_{m[i}\mathring k_{\pm}{}_{j]} \mp \frac{1}{2} \hodge{Y}_{mp}\tilde\pi_\pm{}^p{}_{ij} \\
		=& \mp A^{-1} \partial_{m} A \, \hat \omega_{\pm}{}_{ij} \mp \frac{1}{24} \hodge{S}_{mijpqr}\tilde\pi_\pm{}^{pqr} \pm\frac{1}{4}\hodge{F}_{mijp}k_\pm{}^p \\
		&+ \frac{1}{2}\delta_{m[i}F_{j]p}\mathring k_{\pm}{}^p	+ \frac{3}{4}F_{[mi}\mathring k_{\pm}{}_{j]} \mp \frac{3}{4}\hodge{Y}_{[m|p|} \tilde\pi_\pm{}^p{}_{ij]} \\
		&\pm \frac{1}{4} \delta_{m[i}\hodge{Y}_{|pq|}\tilde\pi_\pm{}^{pq}{}_{j]} \pm \frac{1}{2}X\delta_{m[i}k_\pm{}_{j]}~,
	\end{split}
\end{equation}

\begin{equation}
	\begin{split}
		{\cal D}^{\cal F}_m \mathring \omega_{\pm}{}_{ij} \defeq& \nabla_m \mathring \omega_{\pm}{}_{ij} + H_{m[i|p|}\hat \omega_{\pm}{}^p{}_{j]} \pm \frac{1}{2} \hodge{F}_{m[i|pq|}\tilde\pi_\pm{}^{pq}{}_{j]} \mp \hodge{Y}_{m[i}k_\pm{}_{j]} \\
		=& \mp A^{-1} \partial_{m} A \, \mathring \omega_{\pm}{}_{ij} - \frac{1}{2} S \delta_{m[i}\mathring k_{\pm}{}_{j]} \pm \frac{3}{8} \hodge{F}_{[mi|pq|}\tilde\pi_\pm{}^{pq}{}_{j]} \\
		&\pm \frac{1}{12} \delta_{m[i}\hodge{F}_{j]pqr}\tilde\pi_\pm{}^{pqr} \mp \frac{1}{2} \delta_{m[i}\hodge{Y}_{j]p}k_\pm{}^p \mp \frac{3}{4} \hodge{Y}_{[mi}k_\pm{}_{j]} \\
		&+ \frac{1}{4}Y_{mijp}\mathring k_{\pm}{}^p \pm \frac{1}{4}X\tilde\pi_\pm{}_{mij}~,
	\end{split}
\end{equation}

\begin{equation}
	\begin{split}
		{\cal D}^{\cal F}_m \tilde{\pi}_{\pm}{}_{ijk} \defeq& \nabla_m \tilde{\pi}_{\pm}{}_{ijk} \mp \frac{3}{2} \hodge{H}_{m[ij}\mathring k_{\pm}{}_{k]} - \frac{3}{2}H_{m[ij}k_\pm{}_{k]} \pm \frac{3}{2} \hodge{F}_{m[ij|p|}\mathring \omega_{\pm}{}^p{}_{k]} \\
		&\mp \frac{3}{2}\hodge{Y}_{m[i}\hat \omega_{\pm}{}_{jk]} \\
		=& \mp A^{-1} \partial_{m} A \, \tilde{\pi}_{\pm}{}_{ijk} \mp \frac{3}{2}\delta_{m[i}\hodge{H}_{jk]p}\mathring k_{\pm}{}^p \mp 2 \hodge{H}_{[mij}\mathring k_{\pm}{}_{k]} \mp \frac{1}{8} \hodge{S}_{mijkpq} \hat \omega_{\pm}{}^{pq} \\
		&- \frac{3}{4}\delta_{m[i}F_{jk]}f_\pm \mp \frac{3}{8} \delta_{m[i}\hodge{F}_{jk]pq}\mathring \omega_{\pm}{}^{pq} \pm \hodge{F}_{[mij|p|}\mathring \omega_{\pm}{}^p{}_{k]} \mp \frac{3}{2}\delta_{m[i}\hodge{Y}_{j|p|} \hat \omega_{\pm}{}^p{}_{k]} \\
		&\mp \frac{3}{2}\hodge{Y}_{[mi}\hat \omega_{\pm}{}_{jk]} + \frac{1}{4}Y_{mijk}\tilde{f}_\pm \mp \frac{3}{4}X\delta_{m[i}\mathring \omega_{\pm}{}_{jk]}~,
	\end{split}
\end{equation}
where  $\mathcal{D}^\mathcal{F}$ is the  minimal  connection.

Similarly, a basis in the space of form bilinears on $ M^6$, up to Hodge duality, which are skew-symmetric in the exchange of Killing spinors $\chi_\pm^r$ and $\chi_\pm^s$ is
\begin{equation} \label{AdS4_Skew_Bilinears}
	\begin{gathered}
		\hat f^{rs}_{\pm} = \H{\chi^r_\pm}{\Gamma_{zx}\chi^s_\pm}~, \quad \mathring f^{rs}_{\pm} = \H{\chi^r_\pm}{\Gamma_{zx}\Gamma_{11}\chi^s_\pm}~, \\
		\hat k^{rs}_{\pm} = \H{\chi^r_\pm}{\Gamma_{izx}\chi^s_\pm} \, \e^i~, \quad \tilde k^{rs}_\pm = \H{\chi^r_\pm}{\Gamma_i \Gamma_{11}\chi^s_\pm} \, \e^i~, \\
		\omega^{rs}_\pm = \frac{1}{2} \H{\chi^r_\pm}{\Gamma_{ij}\chi^s_\pm} \, \e^i \wedge \e^j~, \quad \tilde{\omega}^{rs}_\pm = \frac{1}{2} \H{\chi^r_\pm}{\Gamma_{ij}\Gamma_{11}\chi^s_\pm} \, \e^i \wedge \e^j~, \\
		\pi^{rs}_\pm = \frac{1}{3!} \H{\chi^r_\pm}{\Gamma_ {ijk} \chi^s_\pm} \, \e^i \wedge \e^j \wedge \e^k~.
	\end{gathered}
\end{equation}
The associated TCFH is

\begin{equation}
	\begin{split}
		{\cal D}^{\cal F}_m \hat f_{\pm} \defeq& \nabla_m \hat f_{\pm} \\
		=& \mp A^{-1} \partial_{m} A \, \hat f_{\pm} \, - \frac{1}{4} S \hat k_{\pm m} \mp \frac{1}{4!} \hodge{F}_{mpqr} \pi_\pm{}^{pqr} \pm \frac{1}{4} \hodge{Y}_{mp} \tilde k_\pm{}^p~,
	\end{split}
\end{equation}

\begin{equation}
	\begin{split}
		{\cal D}^{\cal F}_m \mathring f_{\pm} \defeq& \nabla_m \mathring f_{\pm} \\
		=& \mp A^{-1} \partial_{m} A \, \mathring f_{\pm} \, - \frac{1}{4} F_{mp} \hat k_{\pm}{}^p \pm \frac{1}{8} \hodge{Y}_{pq} \pi_\pm{}^{pq}{}_m \pm \frac{1}{4} X \tilde k_{\pm m}~,
	\end{split}
\end{equation}

\begin{equation}
	\begin{split}
		{\cal D}^{\cal F}_m \tilde k_\pm{}_i \defeq& \nabla_m \tilde k_\pm{}_i + \frac{1}{4} H_{mpq} \pi_\pm{}^{pq}{}_i - \frac{1}{2} F_{mp} \omega_\pm{}^p{}_i \\
		=& \mp A^{-1} \partial_{m} A \, \tilde{k}_\pm{}_i \, - \frac{1}{4} S \omega_{\pm mi} + \frac{1}{8} \delta_{mi} F_{pq} \omega_\pm{}^{pq} - \frac{1}{2} F_{[m|p|} \omega_\pm{}^p{}_{i]} \\
		&\pm \frac{1}{4} \hodge{Y}_{mi} \hat f_{\pm} - \frac{1}{8} Y_{mipq} \tilde\omega_\pm{}^{pq} \mp \frac{1}{4} X \delta_{mi} \mathring f_{\pm}~,
	\end{split}
\end{equation}

\begin{equation}
	\begin{split}
		{\cal D}^{\cal F}_m \hat k_{\pm}{}_i \defeq& \nabla_m \hat k_{\pm}{}_i \mp \frac{1}{4} \hodge{H}_{mpq} \pi_\pm{}^{pq}{}_i \pm \frac{1}{2} \hodge{Y}_{mp} \tilde\omega_\pm{}^p{}_i \\
		=& \mp A^{-1} \partial_{m} A \, \hat k_{\pm}{}_i \, \mp \frac{1}{12} \delta_{mi} \hodge{H}_{pqr} \pi_\pm{}^{pqr} \mp \frac{1}{2} \hodge{H}_{[m|pq|}\pi_\pm{}^{pq}{}_{i]} \\
		&- \frac{1}{4} S \delta_{mi} \hat f_{\pm} \mp \frac{1}{8} \hodge{F}_{mipq} \omega_\pm{}^{pq} + \frac{1}{4} F_{mi} \mathring f_{\pm} \mp \frac{1}{8} \delta_{mi} \hodge{Y}_{pq} \tilde\omega_\pm{}^{pq} \\
		&\pm \frac{1}{2} \hodge{Y}_{[m|p|}\tilde\omega_\pm{}^p{}_{i]} \pm \frac{1}{4} X \omega_{\pm mi}~,
	\end{split}
\end{equation}

\begin{equation}
	\begin{split}
		{\cal D}^{\cal F}_m \omega_{\pm ij} \defeq& \nabla_m \omega_{\pm ij} + H_{m[i|p|} \tilde\omega_\pm{}^p{}_{j]} + F_{m[i} \tilde k_{\pm j]} + \frac{1}{2} Y_{m[i|pq|} \pi_\pm{}^{pq}{}_{j]} \\
		=& \mp A^{-1} \partial_{m} A \, \omega_{\pm ij} \, - \frac{1}{4} S \pi_{\pm mij} \mp \frac{1}{4} \hodge{F}_{mijp} \hat k_{\pm}{}^p + \frac{1}{2} \delta_{m[i} F_{j]p} \tilde k_\pm{}^p \\
		&+ \frac{3}{4} F_{[mi} \tilde k_{\pm j]} + \frac{1}{12} \delta_{m[i} Y_{j]pqr} \pi_\pm{}^{pqr} + \frac{3}{8} Y_{[mi|pq|} \pi_\pm{}^{pq}{}_{j]} \mp \frac{1}{2} X \delta_{m[i} \hat k_{\pm j]}~,
	\end{split}
\end{equation}

\begin{equation}
	\begin{split}
		{\cal D}^{\cal F}_m \tilde\omega_{\pm ij} \defeq& \nabla_m \tilde\omega_{\pm ij} + H_{m[i|p|} \omega_\pm{}^p{}_{j]} + \frac{1}{2} F_{mp} \pi_\pm{}^p{}_{ij} \pm \hodge{Y}_{m[i} \hat k_{\pm j]} \\
		=& \mp A^{-1} \partial_{m} A \, \tilde\omega_{\pm ij} \, - \frac{1}{2} S \delta_{m[i} \tilde k_{\pm j]} - \frac{1}{4} \delta_{m[i}F_{|pq|} \pi_\pm{}^{pq}{}_{j]} \\
		&+ \frac{3}{4} F_{[m|p|} \pi_\pm{}^p{}_{ij]} \pm \frac{1}{2} \delta_{m[i} \hodge{Y}_{j]p} \hat k_{\pm}{}^p \pm \frac{3}{4} \hodge{Y}_{[mi} \hat k_{\pm j]} \\
		&+ \frac{1}{4} Y_{mijp} \tilde k_\pm{}^p - \frac{1}{4!} \hodge{X}_{mijpqr} \pi_\pm{}^{pqr}~,
	\end{split}
\end{equation}

\begin{equation}
	\begin{split}
		{\cal D}^{\cal F}_m \pi_{\pm ijk} \defeq& \nabla_m \pi_{\pm ijk} \mp \frac{3}{2} \hodge{H}_{m[ij} \hat k_{\pm k]} - \frac{3}{2} H_{m[ij} \tilde k_{\pm k]} - \frac{3}{2} F_{m[i} \tilde \omega_{\pm jk]} - \frac{3}{2} Y_{m[ij|p|} \omega_\pm{}^p{}_{k]} \\
		=& \mp A^{-1} \partial_{m} A \, \pi_{\pm ijk} \, \mp \frac{3}{2} \delta_{m[i} \hodge{H}_{jk]p} \hat k_{\pm}{}^p \mp 2 \hodge{H}_{[mij} \hat k_{\pm k]} - \frac{3}{4} S \delta_{m[i} \omega_{\pm jk]} \\
		&\mp \frac{1}{4} \hodge{F}_{mijk} \hat f_{\pm} - \frac{3}{2} \delta_{m[i} F_{j|p|} \tilde \omega_\pm{}^p{}_{k]} - \frac{3}{2} F_{[mi} \tilde\omega_{\pm jk]} + \frac{3}{8} \delta_{m[i} Y_{jk]pq} \omega_\pm{}^{pq} \\
		&- Y_{[mij|p|} \omega_\pm{}^p{}_{k]} \mp \frac{3}{4} \delta_{m[i} \hodge{Y}_{jk]} \mathring f_{\pm}
+ \frac{1}{8} \hodge{X}_{mijkpq} \tilde\omega_\pm{}^{pq}~,
	\end{split}
\end{equation}
where, again, ${\cal D}^{\cal F}$ is the minimal connection.

The domain that the minimal TCFH connection $\mathcal{D}^\mathcal{F}$ acts factorises into  the space of symmetric form bilinears, \eqref{AdS4_Sym_Bilinears}, and the space of skew-symmetric form bilinears, \eqref{AdS4_Skew_Bilinears} in the exchange of the two Killing spinors $\chi_\pm^r$ and $\chi_\pm^s$. A direct counting of dimensions reveals that the reduced holonomy of $\mathcal{D}^\mathcal{F}$ must be contained in   $GL(64) \times GL(64)$. But as $\mathcal{D}^\mathcal{F}$ acts trivially on the scalars  $f$, $\tilde{f}$, $\hat{f}$ and $\mathring{f}$, its reduced holonomy is contained in    $GL(62) \times GL(62)$.

\section{The TCFH of warped AdS\textsubscript{n}, $n \geq 5$, backgrounds}

\subsection{Fields and Killing spinors}

The bosonic fields of warped AdS$_n$, $AdS_n \times_w M^{10 - n}$, $n \geq 5$, backgrounds with internal space $M^{10-n}$  of (massive) IIA backgrounds can be written  as follows
\bea
		&&g = 2\, \e^+ \e^- + (\e^z)^2+  \sum\limits_{a = 1}^{n - 3} (\e^a)^2   + g(M^{10 - n})~, \cr
&&
		G = G, \quad H = H, \quad F = F, \quad S = m e^\Phi , \quad \Phi = \Phi~,
	\eea
where $g(M^{10-n})$ is a metric on $M^{10-n}$, $m$ is a constant, $\Phi \in C^\infty(M^{10 - n})$, $F \in \Omega^2(M^{10 - n})$, $H \in \Omega^3(M^{10 - n})$ and $G \in \Omega^4(M^{10 - n})$.  For sufficiently large $n$, some of the fluxes may vanish; for example $G$ vanishes for $n \geq 7$. Further,
\begin{equation}
	\begin{gathered}
		\e^+ = \dd u, \quad \e^- = \dd r - \frac{2}{\ell} r \dd z - 2 r A^{-1} \dd A, \quad \e^z=A \dd z~, \quad \e^a=A  e^{z/\ell}  \dd x^a~,
\quad\e^i = e^i{}_J \dd y^J~,
	\end{gathered}
\end{equation}
is a  pseudo-orthonormal frame on $AdS_n \times_w M^{10 - n}$ with $g(M^{10-n})=\delta_{ij} \e^i \e^j$, where $y$ are coordinates on $M^{10-n}$ and  $(u, r, z, x^a)$ are the remaining coordinates of the spacetime. As in previous cases, $A\in C^\infty(M^{10-n})$ is the warp factor and after a coordinate transformation the spacetime metric $g$ can be written in the usual warped form involving the standard  metric on AdS$_n$ of radius $\ell$.

Again the Killing spinors of these backgrounds can be expressed as $\epsilon=\epsilon(u,r, z, x^a, \sigma_\pm, \tau_\pm)$, where $\sigma_\pm$ and $\tau_\pm$ depend only on the coordinates of $M^{10-n}$ and $\Gamma_\pm\sigma_\pm=\Gamma_\pm \tau_\pm=0$ \cite{bgpiia}.  Furthermore, the gravitino KSE along $M^{10-n}$ requires that $\mathcal{D}_m^{(\pm)} \chi_\pm = 0$ with
\begin{equation}
	\begin{split}
			\mathcal{D}_m^{(\pm)} = \nabla_m  \pm & \frac{1}{2} A^{-1} \partial_{m} A  +\frac{1}{8} \slashed{H}_{m}\Gamma_{11} + \frac{1}{8} S \Gamma_{m} \\
			&+ \frac{1}{16} \slashed{F} \Gamma_{m} \Gamma_{11} + \frac{1}{192} \slashed{G} \Gamma_{m}~,
	\end{split}
\end{equation}
where $\nabla_m$, $m=1,\dots, 10-n$, is the spin connection of $g(M^{10 - n})$ and $\chi_\pm$ stands for either $\sigma_\pm$ or $\tau_\pm$.

TCFH of warped AdS$_n$ backgrounds will be stated below for each $n$, $5\leq n\leq 7$.  As the computation is similar to those that have already been described in previous cases, we shall simply state the results.

\subsection{The TCFH of warped AdS\textsubscript{5} backgrounds}

A basis in the space of form bilinears\footnote{As for warped AdS$_4$ backgrounds a more general class of form bilinears can be considered but the choices below for all AdS$_n$, $n\geq 5$, backgrounds will suffice.} on $M^5$, up to Hodge duality, which are symmetric in the exchange of Killing spinors $\chi_\pm^r$ and $\chi_\pm^s$ is
\begin{equation} \label{AdS5_Sym_Bilinears}
	\begin{gathered}
		f^{rs}_\pm = \H{\chi^r_\pm}{\chi^s_\pm}~, \quad 	 \tilde{f}^{rs}_\pm = \H{\chi^r_\pm}{\Gamma_{11} \chi^s_\pm}~, \quad	\mathring f^{rs}_{\pm} = \H{\chi^r_\pm}{\Gamma_ {zx_1x_2}\Gamma_{11} \chi^s_\pm}~, \\
		 k^{rs}_\pm = \H{\chi^r_\pm}{\Gamma_ i\chi^s_\pm} \, \e^i~, \quad \hat k^{rs}_{\pm} = \H{\chi^r_\pm}{\Gamma_{izx_1x_2}\chi^s_\pm} \, \e^{i}~, \\
		\mathring k^{rs}_{\pm} = \H{\chi^r_\pm}{\Gamma_{izx_1x_2}\Gamma_{11}\chi^s_\pm} \, \e^{i}~, \quad \hat \omega^{rs}_{\pm} = \frac{1}{2} \H{\chi^r_\pm}{\Gamma_ {ijzx_1x_2} \chi^s_\pm} \, \e^i \wedge \e^j~.
	\end{gathered}
\end{equation}
The TCFH  is

\begin{equation}
	\begin{split}
		{\cal D}^{\cal F}_m f_\pm \defeq& \nabla_m f_\pm \\
		=& \mp A^{-1} \partial_{m} A \, f_\pm  - \frac{1}{4}Sk_\pm{}_m \pm \frac{1}{8}\hodge{F}_{mpq}\hat \omega_{\pm}{}^{pq} \mp \frac{1}{4}\hodge{G}_{m}\mathring f_{\pm}~,
	\end{split}
\end{equation}

\begin{equation}
	\begin{split}
		{\cal D}^{\cal F}_m \tilde{f}_\pm \defeq& \nabla_m \tilde{f}_\pm \\
		=& \mp A^{-1} \partial_{m} A \, \tilde{f}_\pm - \frac{1}{4} F_{mp}k_\pm{}^p \pm \frac{1}{4}\hodge{G}_p \hat \omega_{\pm}{}^p{}_m~,
	\end{split}
\end{equation}

\begin{equation}
	\begin{split}
		{\cal D}^{\cal F}_m \mathring f_{\pm} \defeq& \nabla_m \mathring f_{\pm} \\
		=& \mp A^{-1} \partial_{m} A \, \mathring f_{\pm} -\frac{1}{4}H_{mpq}\hat \omega_{\pm}{}^{pq} -\frac{1}{4}S\mathring k_{\pm}{}_m + \frac{1}{4} F_{mp}\hat k_{\pm}{}^{p} \mp  \frac{1}{4} \hodge{G}_{m}f_\pm~,
	\end{split}
\end{equation}

\begin{equation}
	\begin{split}
		{\cal D}^{\cal F}_m k_\pm{}_i \defeq& \nabla_m k_\pm{}_i \mp \frac{1}{2}\hodge{H}_{mp}\hat \omega_{\pm}{}^p{}_i \pm \frac{1}{2} \hodge{G}_{m}\mathring k_{\pm}{}_i \\
		=& \mp A^{-1} \partial_{m} A \, k_\pm{}_i \, \mp \hodge{H}_{[m|p|}\hat \omega_{\pm}{}^p{}_{i]} \pm \frac{1}{4}\delta_{mi}\hodge{H}_{pq}\hat \omega_{\pm}{}^{pq} -\frac{1}{4} \delta_{mi} Sf_\pm \\
		&\pm \frac{1}{4} \hodge{F}_{mip}\hat k_{\pm}{}^{p} + \frac{1}{4} F_{mi}\tilde{f}_\pm \pm \frac{1}{4} \delta_{mi}\hodge{G}_{p}\mathring k_{\pm}{}^{p} \pm \frac{1}{2}\hodge{G}_{[m}\mathring k_{\pm}{}_{i]}~,
	\end{split}
\end{equation}

\begin{equation}
	\begin{split}
		{\cal D}^{\cal F}_m \hat k_{\pm}{}_i \defeq& \nabla_m \hat k_{\pm}{}_i \\
		=& \mp A^{-1} \partial_{m} A \, \hat k_{\pm}{}_i \, - \frac{1}{2} H_{mip} \mathring k_{\pm}{}^p - \frac{1}{4}S\hat \omega_{\pm}{}_{mi} \mp \frac{1}{4} \hodge{F}_{mip}k_\pm{}^p \\
		&- \frac{1}{4} F_{mi} \mathring f_{\pm} - \frac{1}{8} G_{mipq} \hat \omega_{\pm}{}^{pq}~,
	\end{split}
\end{equation}

\begin{equation}
	\begin{split}
		{\cal D}^{\cal F}_m \mathring k_{\pm}{}_i \defeq& \nabla_m \mathring k_{\pm}{}_i + \frac{1}{2} F_{mp} \hat \omega_{\pm}{}^p{}_i \pm \frac{1}{2} \hodge{G}_m k_\pm{}_i \\
		=& \mp A^{-1} \partial_{m} A \, \mathring k_{\pm}{}_i \, - \frac{1}{2} H_{mip} \hat k_{\pm}{}^p - \frac{1}{4}\delta_{mi} S\mathring f_{\pm} - \frac{1}{8} \delta_{mi}F_{pq}\hat \omega_{\pm}{}^{pq} \\
		&+  \frac{1}{2} F_{[m|p|} \hat \omega_{\pm}{}^p{}_{i]} \pm \frac{1}{4} \delta_{mi} \hodge{G}_p k_\pm{}^p \pm \frac{1}{2} \hodge{G}_{[m} k_\pm{}_{i]}~,
	\end{split}
\end{equation}

\begin{equation}
	\begin{split}
		{\cal D}^{\cal F}_m \hat \omega_{\pm}{}_{ij} \defeq& \nabla_m \hat \omega_{\pm}{}_{ij} \mp \hodge{H}_{m[i} k_\pm{}_{j]} - F_{m[i} \mathring k_{\pm}{}_{j]} \\
		=& \mp A^{-1} \partial_{m} A \, \hat \omega_{\pm}{}_{ij} \, \mp \delta_{m[i} \hodge{H}_{j]p} k_\pm{}^p \mp \frac{3}{2} \hodge{H}_{[mi} k_\pm{}_{j]} \\
		&+ \frac{1}{2} H_{mij} \mathring f_{\pm} - \frac{1}{2} S \delta_{m[i} \hat k_{\pm}{}_{j]} \pm \frac{1}{4} \hodge{F}_{mij} f_\pm - \frac{1}{2} \delta_{m[i} F_{j]p} \mathring k_{\pm}{}^p \\
		&- \frac{3}{4} F_{[mi} \mathring k_{\pm}{}_{j]} \pm \frac{1}{2} \delta_{m[i} \hodge{G}_{j]} \tilde{f}_\pm + \frac{1}{4} G_{mijp} \hat k_{\pm}{}^p~,
	\end{split}
\end{equation}
where $\nabla$ is the frame connection of $g(M^5)$.

A basis in the space of form bilinears on $M^5$, up to Hodge duality, which are skew-symmetric in the exchange of  $\chi^r$ and $\chi^s$ is
\begin{equation} \label{AdS5_Skew_Bilinears}
	\begin{gathered}
		\hat f^{rs}_{\pm} = \H{\chi^r_\pm}{\Gamma_{zx_1x_2}\chi^s_\pm}~, \quad \tilde k^{rs}_\pm = \H{\chi^r_\pm}{\Gamma_i \Gamma_{11}\chi^s_\pm} \, \e^i~, \\
		\omega^{rs}_\pm = \frac{1}{2} \H{\chi^r_\pm}{\Gamma_{ij}\chi^s_\pm} \, \e^i \wedge \e^j~, \quad \tilde{\omega}^{rs}_\pm = \frac{1}{2} \H{\chi^r_\pm}{\Gamma_{ij}\Gamma_{11}\chi^s_\pm} \, \e^i \wedge \e^j~, \\
		\mathring \omega^{rs}_{\pm} = \frac{1}{2} \H{\chi^r_\pm}{\Gamma_{ijzx_1x_2}\Gamma_{11}\chi^s_\pm} \, \e^i \wedge \e^j~.
	\end{gathered}
\end{equation}
The  TCFH is
\begin{equation}
	\begin{split}
		{\cal D}^{\cal F}_m \hat f_{\pm} \defeq& \nabla_m \hat f_{\pm} \\
		=& \mp A^{-1} \partial_{m} A \, \hat f_{\pm} \, - \frac{1}{4} H_{mpq} \mathring \omega_{\pm}{}^{pq} \mp \frac{1}{8} \hodge{F}_{mpq} \omega_\pm{}^{pq} \pm \frac{1}{4} \hodge{G}_p \tilde\omega_\pm{}^p{}_m~,
	\end{split}
\end{equation}

\begin{equation}
	\begin{split}
		{\cal D}^{\cal F}_m \tilde k_\pm{}_i \defeq& \nabla_m \tilde k_\pm{}_i \mp \frac{1}{2} \hodge{H}_{mp} \mathring \omega_{\pm}{}^p{}_i - \frac{1}{2} F_{mp} \omega_\pm{}^p{}_i \\
		=& \mp A^{-1} \partial_{m} A \, \tilde{k}_\pm{}_i \, \pm \frac{1}{4} \delta_{mi} \hodge{H}_{pq} \mathring \omega_{\pm}{}^{pq} \mp \hodge{H}_{[m|p|} \mathring \omega_{\pm}{}^p{}_{i]} \\
		&- \frac{1}{4} S \tilde\omega_{\pm mi} + \frac{1}{8} \delta_{mi} F_{pq} \omega_\pm{}^{pq} - \frac{1}{2} F_{[m|p|} \omega_\pm{}^p{}_{i]} - \frac{1}{8} G_{mipq} \tilde\omega_\pm{}^{pq}~,
	\end{split}
\end{equation}

\begin{equation}
	\begin{split}
		{\cal D}^{\cal F}_m \omega_\pm{}_{ij} \defeq& \nabla_m \omega_\pm{}_{ij} + H_{m[i|p|} \tilde\omega_\pm{}^p{}_{j]} + F_{m[i} \tilde k_{\pm j]} \pm \frac{1}{2} \hodge{G}_m \mathring \omega_{\pm ij} \\
		=& \mp A^{-1} \partial_{m} A \, \omega_\pm{}_{ij} \, \pm \frac{1}{8} \hodge{S}_{mijpq} \mathring \omega_{\pm}{}^{pq} \mp \frac{1}{4} \hodge{F}_{mij} \hat f_{\pm} + \frac{1}{2} \delta_{m[i} F_{j]p} \tilde k_\pm{}^p \\
		&+ \frac{3}{4} F_{[mi} \tilde k_{\pm j]} \pm \frac{1}{2} \delta_{m[i}\hodge{G}_{|p|} \mathring \omega_{\pm}{}^p{}_{j]} \pm \frac{3}{4} \hodge{G}_{[m}\mathring \omega_{\pm ij]}~,
	\end{split}
\end{equation}

\begin{equation}
	\begin{split}
		{\cal D}^{\cal F}_m \tilde\omega_\pm{}_{ij} \defeq& \nabla_m \tilde\omega_\pm{}_{ij} + H_{m[i|p|}\omega_\pm{}^p{}_{j]} \pm \hodge{F}_{m[i|p|} \mathring \omega_{\pm}{}^p{}_{j]} \\
		=& \mp A^{-1} \partial_{m} A \, \tilde\omega_\pm{}_{ij} - \frac{1}{2} S\delta_{m[i} \tilde k_{\pm j]} \mp \frac{1}{4} \delta_{m[i} \hodge{F}_{j]pq} \mathring \omega_{\pm}{}^{pq} \\
		&\pm \frac{3}{4} \hodge{F}_{[mi|p|} \mathring \omega_{\pm}{}^p{}_{j]} \pm \frac{1}{2} \delta_{m[i} \hodge{G}_{j]} \mathring f_{\pm} + \frac{1}{4} G_{mijp} \tilde k_\pm{}^p~,
	\end{split}
\end{equation}

\begin{equation}
	\begin{split}
		{\cal D}^{\cal F}_m \mathring \omega_{\pm}{}_{ij} \defeq& \nabla_m \mathring \omega_{\pm}{}_{ij} \mp \hodge{H}_{m[i} \tilde k_{\pm j]} \mp \hodge{F}_{m[i|p|} \tilde \omega_\pm{}^p{}_{j]} \pm \frac{1}{2} \hodge{G}_m \omega_{\pm ij} \\
		=& \mp A^{-1} \partial_{m} A \, \mathring \omega_{\pm}{}_{ij} \mp \delta_{m[i} \hodge{H}_{j]p} \tilde k_\pm{}^p \mp \frac{3}{2} \hodge{H}_{[mi} \tilde k_{\pm j]} + \frac{1}{2} H_{mij} \hat f_{\pm} \\
		& \pm \frac{1}{8} \hodge{S}_{mijpq} \omega_\pm{}^{pq} \pm \frac{1}{4} \delta_{m[i} \hodge{F}_{j]pq} \tilde\omega_\pm{}^{pq} \mp \frac{3}{4} \hodge{F}_{[mi|p|}\tilde\omega_\pm{}^p{}_{j]} \\
		& \pm \frac{1}{2} \delta_{m[i} \hodge{G}_{|p|} \omega_\pm{}^p{}_{j]} \pm \frac{3}{4} \hodge{G}_{[m}\omega_{\pm ij]}~.
	\end{split}
\end{equation}
As the domain of the TCFH minimal connection, ${\cal D}^{\cal F}$, factorises on the symmetric and skew-symmetric form bilinears under the exchange of $\chi_\pm^r$ and $\chi_\pm^s$ and after taking into account the details of the action of ${\cal D}^{\cal F}$ on the forms, one concludes that the reduced holonomy of ${\cal D}^{\cal F}$ is included in   $GL(20) \times SO(5) \times GL(35)$.

\subsection{The TCFH of warped AdS\textsubscript{6} backgrounds}

A basis in the space of form bilinears on $ M^4$, up to Hodge duality, which are symmetric in the exchange of $\chi_\pm^r$ and $\chi_\pm^s$ is
\begin{equation} \label{AdS6_Sym_Bilinears}
	\begin{gathered}
		f^{rs}_\pm = \H{\chi^r_\pm}{\chi^s_\pm}~, \quad \tilde{f}^{rs}_\pm = \H{\chi^r_\pm}{\Gamma_{11} \chi^s_\pm} , \quad \hat f^{rs}_{\pm} = \H{\chi^r_\pm}{\Gamma_ {zx_1x_2x_3}\chi^s_\pm}~, \\
		\mathring f^{rs}_{\pm} = \H{\chi^r_\pm}{\Gamma_ {zx_1x_2x_3}\Gamma_{11} \chi^s_\pm} , \quad k^{rs}_\pm = \H{\chi^r_\pm}{\Gamma_ i\chi^s_\pm} \, \e^i~, \quad \hat k^{rs}_{\pm} = \H{\chi^r_\pm}{\Gamma_ {izx_1x_2x_3} \chi^s_\pm} \, \e^i~.
	\end{gathered}
\end{equation}
The TCFH is
\begin{equation}
	\begin{split}
		{\cal D}^{\cal F}_m f_{\pm} \defeq& \nabla_m f_{\pm} \\
		=& \mp A^{-1} \partial_{m} A \, f_\pm  - \frac{1}{4}Sk_\pm{}_m \mp \frac{1}{4}\hodge{F}_{mp}\hat k_{\pm}{}^{p}~,
	\end{split}
\end{equation}

\begin{equation}
	\begin{split}
		{\cal D}^{\cal F}_m \tilde{f}_\pm \defeq& \nabla_m \tilde{f}_\pm \\
		=& \mp A^{-1} \partial_{m} A \, \tilde{f}_\pm - \frac{1}{4} F_{mp}k_\pm{}^p \pm \frac{1}{4}\hodge{G} \hat k_{\pm}{}_m~,
	\end{split}
\end{equation}

\begin{equation}
	\begin{split}
		{\cal D}^{\cal F}_m \hat f_{\pm} \defeq& \nabla_m \hat f_{\pm} \\
		=& \mp A^{-1} \partial_{m} A \, \hat f_{\pm} -\frac{1}{4}S\hat k_{\pm}{}_m \mp \frac{1}{4} \hodge{F}_{mp}k_\pm{}^p~,
	\end{split}
\end{equation}

\begin{equation}
	\begin{split}
		{\cal D}^{\cal F}_m \mathring f_{\pm} \defeq& \nabla_m \mathring f_{\pm} \\
		=& \mp A^{-1} \partial_{m} A \, \mathring f_{\pm} -\frac{1}{4}F_{mp}\hat k_{\pm}{}^p \pm \frac{1}{4} \hodge{G} k_\pm{}_m~,
	\end{split}
\end{equation}

\begin{equation}
	\begin{split}
		{\cal D}^{\cal F}_m  k_\pm{}_i \defeq& \nabla_m k_\pm{}_i \mp \frac{1}{2} \hodge{H}_m \hat k_{\pm}{}_i \\
		=& \mp A^{-1} \partial_{m} A \, k_\pm{}_i \, \mp \frac{1}{2} \delta_{mi} \hodge{H}_p \hat k_{\pm}{}^p \mp \hodge{H}_{[m}\hat k_{\pm}{}_{i]} -\frac{1}{4}\delta_{mi} Sf_\pm \\
		&\mp \frac{1}{4} \hodge{F}_{mi}\hat f_{\pm} + \frac{1}{4} F_{mi}\tilde{f}_\pm \mp \frac{1}{4} \delta_{mi}\hodge{G}\mathring f_{\pm}~,
	\end{split}
\end{equation}

\begin{equation}
	\begin{split}
		{\cal D}^{\cal F}_m \hat k_{\pm}{}_i \defeq& \nabla_m \hat k_{\pm}{}_i \mp \frac{1}{2} \hodge{H}_m k_\pm{}_i \\
		=& \mp A^{-1} \partial_{m} A \, \hat k_{\pm}{}_i \, \mp \frac{1}{2} \delta_{mi}\hodge{H}_p k_\pm{}^p \mp \hodge{H}_{[m}k_\pm{}_{i]} - \frac{1}{4}\delta_{mi} S \hat f_{\pm} \\
		&\mp \frac{1}{4}\hodge{F}_{mi} f_\pm + \frac{1}{4} F_{mi} \mathring f_{\pm} \mp \frac{1}{4} \delta_{mi} \hodge{G} \tilde{f}~,
	\end{split}
\end{equation}
where $\nabla$ is the spin connection of $g(M^4)$.

A basis in the space of form bilinears on $ M^4$, up to Hodge duality, which are skew-symmetric in the exchange of $\chi_\pm^r$ and $\chi_\pm^s$ is
\begin{equation} \label{AdS6_Skew_Bilinears}
	\begin{gathered}
		\tilde{k}^{rs}_{\pm} = \H{\chi^r_\pm}{\Gamma_i \Gamma_{11} \chi^s_\pm} \, \e^i~, \quad \mathring k^{rs}_{\pm} = \H{\chi^r_\pm}{\Gamma_{izx_1x_2x_3}\Gamma_{11}\chi^s_\pm} \, \e^i~, \\
		\omega^{rs}_\pm = \frac{1}{2} \H{\chi^r_\pm}{\Gamma_{ij}\chi^s_\pm} \, \e^i \wedge \e^j~, \quad \tilde{\omega}^{rs}_\pm = \frac{1}{2} \H{\chi^r_\pm}{\Gamma_{ij}\Gamma_{11}\chi^s_\pm} \, \e^i \wedge \e^j~.
	\end{gathered}
\end{equation}
The TCFH is
\begin{equation}
	\begin{split}
		{\cal D}^{\cal F}_m \tilde k_\pm{}_i \defeq& \nabla_m \tilde k_\pm{}_i \pm \frac{1}{2} \hodge{H}_m \mathring k_{\pm i} - \frac{1}{2} F_{mp} \omega_\pm{}^p{}_i \\
		=& \mp A^{-1} \partial_{m} A \, \tilde{k}_\pm{}_i \, \pm \frac{1}{2} \delta_{mi} \hodge{H}_p \mathring k_{\pm}{}^p \pm \hodge{H}_{[m} \mathring k_{\pm i]} - \frac{1}{4} S \omega_{\pm mi} \\
		&+ \frac{1}{8} \delta_{mi} F_{pq} \omega_\pm{}^{pq} - \frac{1}{2} F_{[m|p|} \omega_\pm{}^p{}_{i]} - \frac{1}{8} G_{mipq} \tilde\omega_\pm{}^{pq}~,
	\end{split}
\end{equation}

\begin{equation}
	\begin{split}
		{\cal D}^{\cal F}_m \mathring k_{\pm i} \defeq& \nabla_m \mathring k_{\pm i} \pm \frac{1}{2} \hodge{H}_m \tilde k_{\pm i} \pm \frac{1}{2} \hodge{F}_{mp} \tilde \omega_\pm{}^p{}_i \\
		=& \mp A^{-1} \partial_{m} A \, \mathring k_{\pm i} \, \pm \frac{1}{2} \delta_{mi} \hodge{H}_p \tilde k_{\pm}{}^p \pm \hodge{H}_{[m} \tilde k_{\pm i]} \mp \frac{1}{8} \hodge{S}_{mipq}\omega_\pm{}^{pq} \\
		&\mp \frac{1}{8} \delta_{mi} \hodge{F}_{pq} \tilde \omega_\pm{}^{pq} \pm \frac{1}{2} \hodge{F}_{[m|p|} \tilde\omega_\pm{}^p{}_{i]} \mp \frac{1}{4} \hodge{G} \omega_{\pm mi}~,
	\end{split}
\end{equation}

\begin{equation}
	\begin{split}
		{\cal D}^{\cal F}_m \omega_\pm{}_{ij} \defeq& \nabla_m \omega_\pm{}_{ij} + H_{m[i|p|}\tilde \omega_\pm{}^p{}_{j]} + F_{m[i} \tilde k_{\pm j]} \\
		=& \mp A^{-1} \partial_{m} A \, \omega_\pm{}_{ij} \, \mp \frac{1}{4} \hodge{S}_{mijp} \mathring k_{\pm}{}^p + \frac{1}{2} \delta_{m[i} F_{j]p} \tilde k_\pm{}^p \\
		&+ \frac{3}{4} F_{[mi} \tilde k_{\pm j]} \pm \frac{1}{2} \hodge{G} \delta_{m[i} \mathring k_{\pm j]}~,
	\end{split}
\end{equation}

\begin{equation}
	\begin{split}
		{\cal D}^{\cal F}_m \tilde\omega_\pm{}_{ij} \defeq& \nabla_m \tilde\omega_\pm{}_{ij} + H_{m[i|p|} \omega_\pm{}^p{}_{j]} \pm \hodge{F}_{m[i}\mathring k_{\pm j]} \\
		=& \mp A^{-1} \partial_{m} A \, \tilde\omega_\pm{}_{ij} - \frac{1}{2} S \delta_{m[i} \tilde k_{\pm j]} \pm \frac{1}{2} \delta_{m[i} \hodge{F}_{j]p} \mathring k_{\pm}{}^p \\
		&\pm \frac{3}{4} \hodge{F}_{[mi}\mathring k_{\pm j]} + \frac{1}{4} G_{mijp} \tilde k_\pm{}^p~.
	\end{split}
\end{equation}
Notice that the minimal TCFH connection, ${\cal D}^{\cal F}$, acts on the form bilinears $k_\pm+ \hat k_\pm$ and $k_\pm- \hat k_\pm$ as a connection gauging a scale symmetry of the type $k\pm\hat k\rightarrow s^{\pm 1} (k\pm\hat k)$, $s\in \bR-\{0\}$.  Therefore the reduced holonomy of the minimal TCFH connection, ${\cal D}^{\cal F}$, is included in  $SO(5)\times GL(1)  \times GL(20)$.

\subsection{The TCFH of warped AdS\textsubscript{7} backgrounds}

A basis in the space of form bilinears on $ M^3$, up to Hodge duality, which are symmetric in the exchange of $\chi_\pm^r$ and $\chi_\pm^s$ is
\begin{equation} \label{AdS7_Sym_Bilinears}
	\begin{gathered}
		f^{rs}_\pm = \H{\chi^r_\pm}{\chi^s_\pm}~, \quad \tilde{f}^{rs}_\pm = \H{\chi^r_\pm}{\Gamma_{11} \chi^s_\pm}~, \quad \hat f^{rs}_{\pm} = \H{\chi^r_\pm}{\Gamma_ {zx_1\dots x_4} \chi^s_\pm}~, \\
		k^{rs}_\pm = \H{\chi^r_\pm}{\Gamma_ i\chi^s_\pm} \, \e^i~.
	\end{gathered}
\end{equation}
The TCFH is
\begin{equation}
	\begin{split}
		{\cal D}^{\cal F}_m f_{\pm} \defeq& \nabla_m f_{\pm} \\
		=& \mp A^{-1} \partial_{m} A \, f_\pm  - \frac{1}{4}Sk_\pm{}_m \mp \frac{1}{4}\hodge{F}_{m}\hat f_{\pm}~,
	\end{split}
\end{equation}

\begin{equation}
	\begin{split}
		{\cal D}^{\cal F}_m \tilde{f}_\pm \defeq& \nabla_m \tilde{f}_\pm \\
		=& \mp A^{-1} \partial_{m} A \, \tilde{f}_\pm - \frac{1}{4} F_{mp}k_\pm{}^p~,
	\end{split}
\end{equation}

\begin{equation}
	\begin{split}
		{\cal D}^{\cal F}_m \hat f_{\pm} \defeq& \nabla_m \hat f_{\pm} \\
		=& \mp A^{-1} \partial_{m} A \, \hat f_{\pm} \, \pm \frac{1}{2} \hodge{H} k_\pm{}_m \mp \frac{1}{4} \hodge{F}_m f_\pm~,
	\end{split}
\end{equation}

\begin{equation}
	\begin{split}
		{\cal D}^{\cal F}_m k_\pm{}_i \defeq& \nabla_m k_\pm{}_i \\
		=& \mp A^{-1} \partial_{m} A \, k_\pm{}_i \, \mp \frac{1}{2}\delta_{mi} \hodge{H} \hat f_{\pm} -\frac{1}{4}\delta_{mi} Sf_\pm + \frac{1}{4} F_{mi}\tilde{f}_\pm~,
	\end{split}
\end{equation}
where $\nabla$ is the spin connection of $g(M^3)$.

A basis in the space of form bilinears of $ M^3$, up to Hodge duality, which are skew-symmetric in the exchange of $\chi_\pm^r$ and $\chi_\pm^s$ is
\begin{equation} \label{AdS7_Skew_Bilinears}
	\begin{gathered}
		\mathring f^{rs}_{\pm} = \H{\chi^r_\pm}{\Gamma_{zx_1\dots x_4}\Gamma_{11}\chi^s_\pm} , \quad \tilde{k}^{rs}_{\pm} = \H{\chi^r_\pm}{\Gamma_i \Gamma_{11} \chi^s_\pm} \, \e^i , \\
		\hat k^{rs}_{\pm} = \H{\chi^r_\pm}{\Gamma_{izx_1\dots x_4}\chi^s_\pm} \, \e^i , \quad \mathring k^{rs}_{\pm} = \H{\chi^r_\pm}{\Gamma_{izx_1\dots x_4}\Gamma_{11}\chi^s_\pm} \, \e^i~,
	\end{gathered}
\end{equation}
The  TCFH is

\begin{equation}
	\begin{split}
		{\cal D}^{\cal F}_m \mathring f_{\pm} \defeq& \nabla_m \mathring f_{\pm} \\
		=& \mp A^{-1} \partial_{m} A \, \mathring f_{\pm} \pm \frac{1}{2} \hodge{H} \tilde k_{\pm m} - \frac{1}{4} \mathring k_{\pm m} + \frac{1}{4} F_{mp} \hat k_{\pm}{}^p~,
	\end{split}
\end{equation}

\begin{equation}
	\begin{split}
		{\cal D}^{\cal F}_m \tilde k_{\pm i} \defeq& \nabla_m \tilde k_{\pm i} \mp \frac{1}{2} \hodge{F}_m \mathring k_{\pm i} \\
		=& \mp A^{-1} \partial_{m} A \, \tilde k_{\pm i} \mp \frac{1}{2} \hodge{H} \delta_{mi} \mathring f_{\pm} \mp \frac{1}{4} \hodge{S}_{mip} \hat k_{\pm}{}^p \\
		& \mp \frac{1}{4} \delta_{mi} \hodge{F}_p \mathring k_{\pm}{}^p \mp \frac{1}{2} \hodge{F}_{[m} \mathring k_{\pm i]}~,
	\end{split}
\end{equation}

\begin{equation}
	\begin{split}
		{\cal D}^{\cal F}_m \hat k_{\pm i} \defeq& \nabla_m \hat k_{\pm i} \\
		=& \mp A^{-1} \partial_{m} A \, \hat k_{\pm i} - \frac{1}{2} H_{mip} \mathring k_{\pm}{}^p \pm \frac{1}{4} \hodge{S}_{mip} \tilde k_\pm{}^p - \frac{1}{4} F_{mi} \mathring f_{\pm}~,
	\end{split}
\end{equation}

\begin{equation}
	\begin{split}
		{\cal D}^{\cal F}_m \mathring k_{\pm i} \defeq& \nabla_m \mathring k_{\pm i} \mp \frac{1}{2} \hodge{F}_{m} \tilde k_{\pm i} \\
		=& \mp A^{-1} \partial_{m} A \, \mathring k_{\pm i} - \frac{1}{2} H_{mip} \hat k_{\pm}{}^p - \frac{1}{4} S \delta_{mi} \mathring f_{\pm} \\
		&\mp \frac{1}{4} \delta_{mi} \hodge{F}_p \tilde k_\pm{}^p \mp \frac{1}{2} \hodge{F}_{[m} \tilde k_{\pm i]}~.
	\end{split}
\end{equation}
As in the previous AdS$_6$ case, observe that the the minimal TCFH connection, ${\cal D}^{\cal F}$, acts on $\tilde k\pm \mathring k$ like gauging an additional gauge symmetry. Therefore the reduced holonomy of the minimal TCFH connection, ${\cal D}^{\cal F}$, is included in   $ SO(3) \times SO(3) \times GL(1)$.

\section{Symmetries of probes, AdS backgrounds   and TCFHs}

\subsection{Probes and symmetries}

The dynamics of relativistic and spinning particles propagating on warped AdS backgrounds, AdS$_n\times_w M^{10-n}$, have been investigated in detail in \cite{epbgp}. Here we shall summarise some key properties of the dynamics of spinning particles which are relevant for the examples that we shall present below. As we shall consider examples for which the warp factor is constant, the action of spinning particles propagating on the spacetime factorises to an action on  AdS$_n$
and an action on the internal space $M^{10-n}$. The latter can be written as
\bea
A_M=-{i\over2} \int\, dt\, d\theta\, \gamma_{IJ} Dy^I \partial_t y^J~,
\eea
where $y=y(t,\theta)$ is a worldline superfield, $(t,\theta)$ are the worldline coordinates, $\gamma$ is the internal space metric and $D^2=i\partial_t$.  Of course if $M^{10-n}$ is the product of two or more other manifolds, then the action $A_M$ factorises further into actions associated to each manifold in the product.

It turns out that the infinitesimal variation
\bea
\delta y^I=\epsilon \alpha^I{}_{J_1\dots J_{m-1}} Dy^{J_1}\cdots Dy^{J_{m-1}}~,
\eea
associated with a $m$-form $\alpha$ on $M^{10-n}$ is a (hidden) symmetry of $A_M$, iff $\alpha$ is a (standard) KY form, where $\epsilon$ is an infinitesimal parameter.
Below we shall present several examples of IIA AdS backgrounds  where KY forms arise as a consequence of the  TCFH  on their internal spaces. In this way, we shall provide a link between TCFHs and conservation laws of probes propagating on such backgrounds.

\subsection{Examples of TCFH and KY forms}

There are many  IIA AdS backgrounds that we can consider, see e.g. \cite{FR, romans, jose, cvetic, lust1, lust2, passias1, rosa, jggpx}.  As the aim is to provide some examples of backgrounds for which  the TCFHs give rise to symmetries for spinning particle probes, we shall not be comprehensive and instead focus on AdS backgrounds that arise as near horizon geometries of intersecting  branes \cite{gppt, aat, jgkt}, see also \cite{bps}. In the analysis that follows, we shall present a ansatz which includes  the near horizon geometry of intersecting branes under consideration and proceed to demonstrate that the associated TCFH gives rise to KY forms on the internal space. In turn these generate symmetries for spinning particle probes and so demonstrate a relation  between TCFHs and probe symmetries.

The formulae for the reduced  field equations and KSEs on the internal space of a warped AdS background that we shall use to construct the AdS solutions suitable for our purposes can be found in \cite{bgpiia}.  As it has already been mentioned, these have been  obtained after suitably solving the field equations and KSEs of the theory over the AdS subspace and identifying the remaining equations on the internal space of these backgrounds. Here we shall typically  quote  the  relevant parts of these equations -- for the derivation and the  full  expressions of these equations the reader should consult the original reference.

\subsubsection{An AdS$_3$ solution from a fundamental string on a NS5 brane}

An example of an AdS$_3$ solution  is that which arises as the near horizon geometry of a fundamental string on a NS5-brane background. This configuration has played a prominent role in a microscopic string theory counting of entropy for extreme black holes \cite{strominger, callan}.  A ansatz which includes such a solution is
\bea
g = g_\ell (AdS_3) + g (\bR^4) + g (S^3)~,~~~
		H = p\, \textrm{dvol}_\ell(AdS_3) + q\, \mathrm{dvol}(S^3)~,
	\label{F1NS5}
\eea
 the dilaton is constant, $\Phi = \textrm{const}$, and the rest of the fields are set to zero,  where $g_\ell (AdS_3)$ ($g(S^3)$) and $\textrm{dvol}_\ell(AdS_3)$ ($\textrm{dvol}(S^3)$) are the standard metric and associated volume form of AdS$_3$ ($S^3$) of radius $\ell$ (unit radius), respectively, $g(\bR^4)$ is the Euclidean metric of $\bR^4$ and $p, q \in \mathbb{R}$. From here on we shall adopt the same conventions for the AdS$_n$  $(S^k$) metric and volume form in all the examples below -- $g(\bR^m)$ will always denote the Euclidean metric on $\bR^m$.  Note that $\bR^4$ can be replaced with any Ricci flat manifold, like for example $K_3$, but the choice of $\bR^4$ suffices for the purpose of this example. Moreover as the warp factor $A$ is constant and the radius $\ell$ of AdS$_3$  has been kept arbitrary, so without loss of generality, we have set $A = 1$. Furthermore, the radius of $S^3$ has been set to 1 after possibly an overall rescaling of the spacetime metric and $H$.

To find a solution based on the ansatz (\ref{F1NS5}), one has to determine $p,q$ and $\ell$ after solving the field and KSEs on the $\bR^4\times S^3$ internal space.
As the IIA 4-form flux vanishes, one has that $X = Y = 0$.  Moreover a direct comparison of \eqref{ads3back} with \eqref{F1NS5} reveals that $p = W$ and $Z=q\, \textrm{dvol}(S^3)$ .

To determine the remaining constants $q$ and $\ell$, one first considers the field equation of the dilaton $\Phi$,
\begin{equation}
	\nabla^2 \Phi = - \frac{1}{12} Z^2 + \frac{1}{2} W^2 \equiv 0~,
\end{equation}
which implies that $q^2 = W^2=p^2$. Next, the Einstein field equations along the $S^3$ directions and the field equation of the warp factor
\begin{equation}
	\begin{gathered}
		R_{\alpha \beta}^{S^3} = \frac{1}{4} Z_{\alpha \gamma \delta}Z^{\gamma \delta}{}_{\beta} \equiv 2\, \delta_{\alpha \beta} \,, \\
	\nabla^2 \log A = - \frac{2}{\ell^2} + \frac{1}{2} W^2 \equiv 0~,
	\end{gathered}
\end{equation}
respectively yield $p^2=W^2 = 4$ and $\ell = 1$, i.e. the AdS$_3$ and $S^3$ subspaces have the same radius and $p,q=\pm2$.

Turning attention to the KSEs, and focusing for simplicity on those on $\sigma_+$, the dilatino KSE, $\mathcal{A}^{(+)} \sigma_+ = 0$, with
\begin{equation}
	\mathcal{A}^{(+)} = \frac{1}{12} \slashed{Z} \Gamma_{11} - \frac{1}{2} W \Gamma_z \Gamma_{11}~,
\end{equation}
gives the condition $\Gamma_{(3)} \Gamma_z \sigma_+ = -\sigma_+$ provided we choose\footnote{The treatment of $p=-q$ case follows from that of $p=q$ in a straightforward manner.} $p=q$, where $\Gamma_{(3)}$ is the product of the three gamma matrices along the orthonormal directions tangent to the three sphere. The additional algebraic KSE, , $\Xi_+ \sigma_+ = 0$, which can be found in \cite{{bgpiia}} with
\begin{equation}
	\Xi_+ = - \frac{1}{2 \ell} + \frac{1}{4} W \Gamma_{11}~,
\end{equation}
that arises from the integration of the gravitino KSE along the $z$ directions, results in the condition $\Gamma_{11} \sigma_+ = \sigma_+$, where we have chosen $p=2$. Therefore, we find that $\sigma_+ $ is a spacetime chiral spinor. The solution with $p=-2$ can be investigated in a similar way to that for $p=2$.

The gravitino KSE \eqref{supconeads3} along $\bR^4$ shows that the Killing spinors $\sigma_+$ satisfy the condition $\nabla_i^{\bR^4} \sigma_+ = 0$ and so do not depend on the coordinates of $\bR^4$. Furthermore, the gravitino KSE along  $S^3$  can be written as:
\begin{equation}
	\nabla_\alpha^{S^3} \sigma_+ + \frac{1}{2} \Gamma_\alpha \Gamma_z \sigma_+ = 0~,
\label{f1ns5s3}
\end{equation}
where  we have made use of the conditions $\Gamma_{(3)} \Gamma_z \sigma_+ = -\sigma_+$ and $\Gamma_{11} \sigma_+ = \sigma_+$. This does not impose further constraints on $\sigma_+$. Therefore the only conditions on $\sigma_+$ are $\Gamma_{(3)} \Gamma_z \sigma_+ = -\sigma_+$ and $\Gamma_{11} \sigma_+ = \sigma_+$ and so $\sigma_+$  has  4 independent components.  A similar analysis of the KSEs on $\sigma_-$ and $\tau_\pm$ spinors  yields another 12 independent Killing spinors and so the solution preserves 1/2 of supersymmetry as expected. Note that if $\bR^4$ is replaced by $K_3$ or any other 4-dimensional hyper-K\"ahler manifold $Q^4$ and the orientation of $Q^4$ is chosen to be compatible with the conditions $\Gamma_{(3)} \Gamma_z \sigma_+ = -\sigma_+$ and $\Gamma_{11} \sigma_+ = \sigma_+$, the solution will again preserve 1/2 of supersymmetry. The spinors $\sigma_\pm$ and $\tau_\pm$ will be covariantly constant with respect to the spin connection of the hyper-K\"ahler metric on $X^4$.

A consequence of (\ref{f1ns5s3}) is that the bilinears
\bea
(k^\pm)_\alpha^{rs} =\langle \sigma^r_\pm, \Gamma_\alpha \sigma^s_\pm\rangle~, ~~~(\omega^\pm)_{\alpha\beta}^{rs} =\langle \sigma^r_\pm, \Gamma_{\alpha\beta} \sigma^s_\pm\rangle~,~~~(\varphi^\pm)_{\alpha\beta\gamma}^{rs} =\langle \sigma^r_\pm, \Gamma_{\alpha\beta\gamma} \sigma^s_\pm\rangle~,
\label{s3bi0}
\eea
are CCKY forms on $S^3$, while the bilinears
\bea
(\tilde k^\pm)^\alpha_{rs} =\langle \sigma^r_\pm, \Gamma_\alpha \Gamma_z \sigma^s_\pm\rangle~, ~~~(\tilde \omega^\pm)_{\alpha\beta}^{rs} =\langle, \sigma^r_\pm, \Gamma_{\alpha\beta}\Gamma_z \sigma^s_\pm\rangle~,~~~(\tilde\varphi^\pm)_{\alpha\beta\gamma}^{rs} =\langle \sigma^r_\pm, \Gamma_{\alpha\beta\gamma} \Gamma_z\sigma^s_\pm\rangle~,
\label{s3bi}
\eea
are  KY  forms on $S^3$.  The latter generate symmetries for spinning particle actions on $S^3$.

\subsubsection{An AdS$_2$ solution from intersecting  D2- and D4-branes}

A ansatz which includes the near horizon geometry of two D2- and two D4-branes intersecting on a 0-brane is
\bea
&&g=g_\ell(AdS_2)+ g(S^2)+ g(\bR^2)+ g(\bR^4)~,
\cr
&&
G= \textrm{dvol}_\ell(AdS_2)\wedge \alpha+ \textrm{dvol}(S^2)\wedge \beta~,
\label{d2d2d4d4}
\eea
with constant dilaton $\Phi$ and all other remaining fields set to zero, where  $\ell$ is the radius of AdS$_2$ and $\alpha$ and $\beta$ are constant 2-forms on $\bR^4$.

Assuming that $\bR^4=\bR\langle(\e_3,\e_4, \e_5, \e_6)\rangle$, there is an $SO(4)$ transformation such that the form $\alpha$ can be written as $\alpha=p\, \e^3\wedge \e^4+ q\, \e^5\wedge \e^6$. The isotropy group $SO(2)\times SO(2)$ of $\alpha$ can then be used to choose  $\beta$ without loss of generality as
\bea
\beta=r\, \e^3\wedge \e^4+ s\, \e^5\wedge \e^6+ a\, \e^3\wedge \e^5+ b\, \e^4\wedge \e^6+ c\, \e^4\wedge \e^5~,
\label{betaeqn}
\eea
where all components of $\alpha$ and $\beta$ are constants in $\bR$.

The Einstein equations along $\bR^4$ (with the two indices distinct) imply that $c r=c s=c b=c a=0$. Thus if $c\not=0$, $r=s=b=a=0$. Then the remaining Einstein equations along $\bR^4$ give that $p=q=0$.  Finally, the dilatino KSE  for the ansatz (\ref{d2d2d4d4}) is
\bea
(-{1\over8}\slashed{X}+{1\over 4\cdot 4!}\slashed{Y})\eta_+=0~,
\eea
and gives $c=0$. Therefore all fluxes vanish for this case, so to proceed we take $c=0$.

Setting $c=0$, the dilatino KSE as well as the gravitino KSE along $\bR^4$ can be written for the fluxes (\ref{d2d2d4d4}) as
\bea
&&\big(-p + q I_1+ \Gamma_{(2)} (-r + s I_1)-a I_2-b I_1 I_2\big)\eta_+=0~,
\cr
&& \big(-p + q I_1+ \Gamma_{(2)} (-r + s I_1)-a I_2-b I_1 I_2\big)\Gamma_\mu \eta_+=0~,~~~\mu=3,4,5,6
\eea
where $I_1=\Gamma_{3456}$, $I_2=\Gamma_{(2)} \Gamma_{45}$, $\Gamma_{(2)}$ is the product of two gamma matrices along orthonormal directions tangent to $S^2$ and we have taken $\eta_+$ to be constant along $\bR^4$. Separating the Hertmitian and anti-Hermitian components of the above equations and using that $I_1 \Gamma_\mu=-\Gamma_\mu I_1$ as well as the commutation relations of $\Gamma_\mu$ with $I_2$, one finds that $r,s=0$ and
\bea
(q I_1+p)\eta_+=0~,~~~(b I_1-a)\eta_+=0~,~~~(a I_2+p)\eta_+=0~.
\eea
These can be solved by restricting $\eta_+$ to the eigenspaces of $I_1$ and $I_2$. In turn, one finds that $p,q,a,b$ are proportional to each other with proportionality factor of a sign.  Therefore in all cases, $a^2=b^2=p^2=q^2$. A similar analysis holds for the $\eta_-$ Killing spinors.  As each eigenspace of $I_1$ and $I_2$ on either $\eta_+$ or $\eta_-$ has dimension 4, there are 8 Killing spinors that solve the above KSEs.

After using that $S^2$ has radius 1, the Einstein equation along $S^2$ reveals that $a^2=1$.  In turn the field equation for the warp factor $A$ gives $\ell=1$. Therefore AdS$_2$ and $S^2$ have the same radius. All the remaining field equations are satisfied.

As the gravitino KSE along $\bR^2$ is satisfied, it remains to explore the gravito KSE along $S^2$. This can be written as
\bea
\nabla_\alpha^{S^2}\eta_++{p\over2} \Gamma_{34}\Gamma_\alpha \eta_+=0~.
\eea
This does not impose any additional conditions on $\eta_+$ and the same applies for the corresponding equation on $\eta_-$.  Therefore the solution preserves $1/4$ of supersymmetry.
It follows from this that the 1- and 2-form bilinears along $S^2$ and their duals are either KY or CCKY forms. There are several KY forms.  For example, one can easily show that
$(k^{(\pm)})^{rs}_\alpha= \langle \eta^r_\pm, \Gamma_\alpha \eta^s_\pm\rangle$ and  $(\tilde k^{(\pm)})^{rs}_\alpha= \langle \eta^r_\pm, \Gamma_\alpha \Gamma_{12} \eta^s_\pm\rangle$ are KY forms. The KY forms generate symmetries for spinning
particles propagating on the internal space of these backgrounds.

The background can be generalised somewhat by replacing $\bR^4$ with any other 4-dimensional hyper-K\"ahler manifold $Q^4$. In such a case, $X$ and $Y$ are chosen as
\bea
X=p^r \lambda_r~,~~~Y=\textrm{dvol}(S^2)\wedge  a^r \lambda_r~,
\eea
 where $\lambda$ are the 3 K\"ahler forms of $Q^4$ associated with the hyper-complex structure and $p^r$ and $a^r$ are constant 3-vectors.  Under a frame $SO(4)$ rotation both $p^r$ and $a^r$ transform as $SO(3)$ vectors.  Moreover, the field equation for the magnetic component of the 3-form field strength implies that
 $\delta_{rs} p^r a^s=0$, i.e. they are orthogonal. In such a case, there is an $SO(4)$ rotation such that $p^r \lambda_r=\alpha$ with $p^2=q^2$ and $a^r\lambda_r=\beta$ as in (\ref{betaeqn}) with $r=s=c=0$ and $a^2=b^2$. Moreover the relative signs in the equalities $p=\pm q$ and $a=\pm b$ should be chosen such that $\alpha$ and $\beta$ have the same self-duality properties on $Q^4$. After that the previous analysis on $\bR^4$ can be repeated to solve both KSEs and field equations yielding a new solution preserving again $1/4$ of supersymmetry. The identification of the KY forms on $S^2$ can be done as for $Q^4=\bR^4$.

 \subsubsection{AdS$_3$ solutions from intersecting D2- and D4-branes}

A ansatz that includes   the near horizon geometry AdS$_2$ of a D2- and a D4-brane intersecting on a 1-brane is
\bea
g=g_\ell(AdS_3)+ g(S^3)+ g(\bR^4)~,~~~
G= \textrm{dvol}_\ell(AdS_3)\wedge \alpha+ \textrm{dvol}(S^3)\wedge \beta~,
\label{d2d4}
\eea
with constant dilaton $\Phi$ and all other remaining fields set to zero, where  $\ell$ is the radius of AdS$_3$ and $\alpha$ and $\beta$ are constant 1-forms on $\bR^4$.

First notice that the field equation for the magnetic component of the NS 3-form implies that $\alpha\wedge \beta=0$ and so $\alpha$ and $\beta$ are co-linear, i.e. they are proportional and so write $\beta= p\alpha$. Next the dilatino KSE on $\sigma_+$ and the algebraic KSE $\Xi_+\sigma_+=0$  imply that
\bea
\big(\Gamma_{(3)} \Gamma_z+{1\over p}\big)\sigma_+=0~,~~~\big({1\over\ell}+{\slashed \alpha}\big)\sigma_+=0~,
\label{proads3}
\eea
where $\Gamma_{(3)}$ is the product of three gamma matrices along orthonormal tangent directions of $S^3$, i.e. the Clifford algebra element associated to
$\textrm{dvol}_\ell(AdS_3)$.
The dilaton field equation gives $p=\pm 1$ and so  $\alpha^2=\beta^2$. Moreover the warp factor field equation yields $\alpha^2=4 \ell^{-2}$.

Turning to the Einstein equation along $S^3$, one finds that
\bea
R^{S^3}_{\alpha\beta}={2\over\ell^2} \delta_{\alpha\beta}~.
\eea
As $S^3$ has unit radius, one concludes that $\ell=1$ and so $\alpha^2=\beta^2=4$.  Therefore AdS$_3$ and $S^3$ have the same radius. Furthermore, one can verify that all the remaining field equations and KSEs are satisfied apart from the gravitino KSE along $S^3$.  This can be written using  (\ref{proads3}) as
\bea
\big(\nabla^{S^3}_\gamma+{1\over4} \Gamma_z \slashed{\alpha} \Gamma_\gamma\big)\sigma_+=0~,
\eea
 and gives no additional conditions on $\sigma_+$.  A similar analysis holds for the remaining Killing spinors $\sigma_-$ and $\tau_\pm$. As a result, the solution preserves 1/2 of supersymmetry.

To proceed one can consider the bilinears as in (\ref{s3bi}) and (\ref{s3bi0})  and proceed to demonstrate that these and their Hodge duals on $S^3$ are either KY or CCKY forms.  The former generate symmetries for spinning probes on $S^3$. In particular $k^{(\pm)}$, $\star\omega^{(\pm)}$ and $\varphi^{(\pm)}$ are KY forms on $S^3$.

\subsubsection{AdS$_2$ solutions  from intersecting  D2-branes and fundamental strings}

A ansatz that includes the near horizon geometry of two D2-branes and a fundamental string intersecting on a 0-brane is
\bea
&&g=g_\ell(AdS_2)+g(S^3)+ g(\bR^5)~,
\cr
&&G=\textrm{dvol}_\ell(AdS_2)\wedge X~,~~~H=\textrm{dvol}_\ell(AdS_2)\wedge W~,
\eea
with constant dilaton $\Phi$ and all other remaining fields set to zero, where $X$ and $W$ are a 2-form and 1-form on $\bR^5$, respectively.

The field equation for the magnetic part of the 2-form field strength implies that $i_W X=0$. The dilaton field equation gives $W^2=1/4\,\, X^2$ and the warp factor field equations can be expressed as $W^2=\ell^{-2}$.

Taking $\bR^5=\bR\langle(\e_1,\e_2, \dots, \e_5)\rangle$, there is a $SO(5)$ transformation, up to a possible relabelling of the basis, such that $X=\lambda_1\, \e^1\wedge \e^2+ \lambda_2\, \e^3\wedge \e^4$ and $\lambda_1, \lambda_2\in \bR$. Next if either $\lambda_1$ or  $\lambda_2$ vanish together with $i_W X=0$, one can show that the gravitino KSE on $\eta_+$ along $\bR^5$ becomes inconsistent. Therefore from now on, we take $\lambda_1, \lambda_2\not=0$ and as $i_W X=0$, we have $W=p\, \e^5$.
Using this, the dilatino KSE yields
\bea
\big({1\over2} \lambda_1-{1\over2} \lambda_2 \Gamma_{1234}+ p \Gamma_{12} \Gamma_{11}  \Gamma_5\big)\eta_+=0~.
\eea
 This together with the gravitino KSE along $\bR^5$ imply that
 \bea
 (\lambda_2 \Gamma_{1234}+\lambda_1)\eta_+=0~,~~~( p \Gamma_{12} \Gamma_{11} \Gamma_5+\lambda_1)\eta_+=0~.
 \label{d2d2f1pro}
 \eea
 As a result $\lambda_1^2=\lambda_2^2=p^2=W^2$.

 Restricting the Einstein equation along $S^3$, which has unit radius,  yields $\lambda_1^2=4$. The warp factor field equation in turn gives $\ell=1/2$. Therefore the AdS$_2$ subspace has half the radius of the internal space $S^3$. It remains to explore the gravitino KSE along $S^3$. This can be rewritten as
 \bea
 \big(\nabla_\alpha^{S^3}+{1\over4} \lambda_1 \Gamma_{12} \Gamma_\alpha\big)\eta_+=0~.
 \label{d2d2f1grav}
 \eea
 This does not impose any additional conditions on $\eta_+$. A similar analysis can be carried out for the $\eta_-$ Killing spinors. As a result the solution preserves 1/4 of supersymmetry as a consequence of the conditions (\ref{d2d2f1pro}) on $\eta_+$ and the analogous conditions on $\eta_-$.

 There are several form bilinears that one can consider on $S^3$ like for example those in (\ref{s3bi}) and (\ref{s3bi0}) and their duals on $S^3$. All of them are either   KY or CCKY as a consequence of (\ref{d2d2f1grav}). In particular $k^{(\pm)}$, $\star\omega^{(\pm)}$ and $\varphi^{(\pm)}$ are KY forms   and so generate symmetries   for spinning particles propagating on $S^3$.

 \section{Concluding Remarks}

 We have presented all the TCFHs of massive IIA warped AdS backgrounds. In particular we have shown that the form bilinears of supersymmetric AdS backgrounds satisfy a generalisation of CKY equation with respect to the TCFH connection. In addition we have explored some of the properties of the minimal TCFH connection like its reduced holonomy. Furthermore we have investigated the question on whether the TCFHs give rise to hidden symmetries for probes propagating on the internal space of AdS backgrounds. For this we presented some examples of AdS backgrounds, namely those arising as near horizon geometries of intersecting IIA branes,  and  demonstrated that some of their form bilinears are KY forms. As a result they generate symmetries for spinning particles propagating on the internal space of such backgrounds.
 This work,   together with those in \cite{epbgp, lggp}, completes the investigation of TCFHs of all warped AdS backgrounds of type II theories in 10 and 11 dimensions.

 The extent of the interplay between TCFHs and symmetries of probes propagating on supersymmetric background remains open.  There are certainly many examples of backgrounds that the TCFH conditions  coincide with those required for the invariance  of probe actions  under transformations generated by the form bilinears. For example in the heterotic and common sector cases,  all form bilinears generate symmetries for certain string and particle probes. However for generic type II theories, the relation between TCFH and probe symmetries can only be revealed on a case by case basis after exploring separately the geometric properties of each background. The difficulties lie both in the lack of classification of supersymmetric backgrounds in type II theories and the plethora of probes \cite{coles} that one can consider. A more systematic investigation will require developments both in the understanding the supersymmetric backgrounds of type II theories as well a better handle on probe actions and their symmetries.

 \section*{Acknowledgments}

JP is supported by the EPSRC studentship grant EP/R513064/1.

\newpage

% \appendix

%\newpage\appendix\section{Conventions}


\newpage

\begin{thebibliography}{99}

\bibitem{gptcfh}
	G.~Papadopoulos,
	``Twisted form hierarchies, Killing-Yano equations and supersymmetric backgrounds,''
	JHEP \textbf{07} (2020), 025
	doi:10.1007/JHEP07(2020)025
	[arXiv:2001.07423 [hep-th]].


%\cite{jgg}
\bibitem{jggp}
J.~Gutowski and G.~Papadopoulos,
``Eigenvalue estimates for multi-form modified Dirac operators,''
J. Geom. Phys. \textbf{160} (2021), 103954
doi:10.1016/j.geomphys.2020.103954
[arXiv:1911.02281 [math.DG]].
%6 citations counted in INSPIRE as of 30 Dec 2022




  \bibitem{carter-b}
  B.~Carter,
  ``Global structure of the Kerr family of gravitational fields,''
  Phys.\ Rev.\  {\bf 174} (1968) 1559.



  \bibitem{penrose}
R. Penrose, Ann. N.Y. Acad. Sci. 224 (1973) 125.


\bibitem{floyd}
  R. Floyd, The dynamics of Kerr fields,. Ph. D. Thesis, London (1973).








  \bibitem{chandrasekhar}
  S.~Chandrasekhar,
  ``The Solution Of Dirac's Equation In Kerr Geometry,''
  Proc.\ Roy.\ Soc.\ Lond.\  A {\bf 349} (1976) 571.

   \bibitem{carter-a}
  B.~Carter,
  ``Killing Tensor Quantum Numbers And Conserved Currents In Curved Space,''
  Phys.\ Rev.\  D {\bf 16} (1977) 3395.

  \bibitem{carter-c}
  B.~Carter and R.~G.~Mclenaghan,
  ``Generalized Total Angular Momentum Operator For The Dirac Equation In
  Curved Space-Time,''
  Phys.\ Rev.\  D {\bf 19} (1979) 1093.

  \bibitem{page}
  P.~Krtous, D.~Kubiznak, D.~N.~Page and V.~P.~Frolov,
  ``Killing-Yano tensors, rank-2 Killing tensors, and conserved quantities in
  higher dimensions,''
  JHEP {\bf 0702} (2007) 004
  [arXiv:hep-th/0612029].

   \bibitem{sfetsos}
  F.~De Jonghe, K.~Peeters and K.~Sfetsos,
  ``Killing-Yano supersymmetry in string theory,''
  Class.\ Quant.\ Grav.\  {\bf 14} (1997) 35
  [arXiv:hep-th/9607203].

\bibitem{lun}
  Y.~Chervonyi and O.~Lunin,
  ``Killing(-Yano) Tensors in String Theory,''
  JHEP {\bf 1509} (2015) 182
  doi:10.1007/JHEP09(2015)182
  [arXiv:1505.06154 [hep-th]].







  \bibitem{revky}
  M.~Cariglia,
  ``Hidden Symmetries of Dynamics in Classical and Quantum Physics,''
  Rev.\ Mod.\ Phys.\  {\bf 86} (2014) 1283
  doi:10.1103/RevModPhys.86.1283
  [arXiv:1411.1262 [math-ph]].


\bibitem{frolov}
V.~Frolov, P.~Krtous and D.~Kubiznak,
``Black holes, hidden symmetries, and complete integrability,''
Living Rev. Rel. \textbf{20} (2017) no.1, 6
doi:10.1007/s41114-017-0009-9
[arXiv:1705.05482 [gr-qc]].





%\cite{Kastor:2004jk}
\bibitem{kt}
D.~Kastor and J.~Traschen,
``Conserved gravitational charges from Yano tensors,''
JHEP \textbf{08} (2004), 045
doi:10.1088/1126-6708/2004/08/045
[arXiv:hep-th/0406052 [hep-th]].
%31 citations counted in INSPIRE as of 08 Jun 2022



%\cite{Papadopoulos:2007gf}
\bibitem{gpky}
G.~Papadopoulos,
``Killing-Yano equations and G-structures,''
Class. Quant. Grav. \textbf{25} (2008), 105016
doi:10.1088/0264-9381/25/10/105016
[arXiv:0712.0542 [hep-th]].
%24 citations counted in INSPIRE as of 08 Jun 2022


%\cite{Howe:2018lwu}
\bibitem{hl}
P.~S.~Howe and U.~Lindstr\"om,
``Some remarks on (super)-conformal Killing-Yano tensors,''
JHEP \textbf{11} (2018), 049
doi:10.1007/JHEP11(2018)049
[arXiv:1808.00583 [hep-th]].
%11 citations counted in INSPIRE as of 08 Jun 2022

%\cite{Lindstrom:2021qrk}
\bibitem{ls1}
U.~Lindstr\"om and \"O.~Sar\i{}o\u{g}lu,
``New currents with Killing\textendash{}Yano tensors,''
Class. Quant. Grav. \textbf{38} (2021) no.19, 195011
doi:10.1088/1361-6382/ac1871
[arXiv:2104.12451 [hep-th]].
%3 citations counted in INSPIRE as of 08 Jun 2022

%\cite{ls}
\bibitem{ls2}
U.~Lindstr\"om and \"O.~Sar\i{}o\u{g}lu,
``Tensionless strings and Killing(-Yano) tensors,''
Phys. Lett. B \textbf{829} (2022), 137088
doi:10.1016/j.physletb.2022.137088
[arXiv:2202.06542 [hep-th]].
%1 citations counted in INSPIRE as of 08 Jun 2022










 \bibitem{gibbons}
  G.~W.~Gibbons, R.~H.~Rietdijk and J.~W.~van Holten,
  ``SUSY in the sky,''
  Nucl.\ Phys.\  B {\bf 404} (1993) 42
  [arXiv:hep-th/9303112].


\bibitem{bvh}
L.~Brink, P.~Di Vecchia and P.~S.~Howe,
``A Lagrangian Formulation of the Classical and Quantum Dynamics of Spinning Particles,''
Nucl. Phys. B \textbf{118} (1977), 76-94
doi:10.1016/0550-3213(77)90364-9








%\cite{Papadopoulos:2022lgb}
\bibitem{epbgp1}
G.~Papadopoulos and E.~P\'erez-Bola\~nos,
``TCFHs, hidden symmetries and M-theory backgrounds,''
Class. Quant. Grav. \textbf{39} (2022) no.24, 245015
doi:10.1088/1361-6382/aca1a2
[arXiv:2201.11563 [hep-th]].
%3 citations counted in INSPIRE as of 30 Dec 2022

%\cite{Grimanellis:2021zxj}
\bibitem{lggpjp}
L.~Grimanellis, G.~Papadopoulos and J.~Phillips,
``TCFHs, hidden symmetries and type II theories,''
JHEP \textbf{07} (2022), 097
doi:10.1007/JHEP07(2022)097
[arXiv:2111.15405 [hep-th]].
%3 citations counted in INSPIRE as of 30 Dec 2022


%\cite{Papadopoulos:2021cgc}
\bibitem{epbgp2}
G.~Papadopoulos and E.~P\'erez-Bola\~nos,
``Symmetries, spinning particles and the TCFH of D=4,5 minimal supergravities,''
Phys. Lett. B \textbf{819} (2021), 136441
doi:10.1016/j.physletb.2021.136441
[arXiv:2101.10709 [hep-th]].
%4 citations counted in INSPIRE as of 30 Dec 2022

\bibitem{epbgp}
G.~Papadopoulos and E.~P\'erez-Bola\~nos,
``The TCFHs of D=11 AdS backgrounds and hidden symmetries,''
[arXiv:2206.04369 [hep-th]].


%\cite{}
\bibitem{lggp}
L.~Grimanellis and G.~Papadopoulos,
``TCFHs, IIB warped AdS backgrounds and hidden symmetries,''
[arXiv:2207.04431 [hep-th]].
%0 citations counted in INSPIRE as of 30 Dec 2022



%\cite{Romans:1985tz}
\bibitem{romans}
L.~J.~Romans,
``Massive N=2a Supergravity in Ten-Dimensions,''
Phys. Lett. B \textbf{169} (1986), 374
doi:10.1016/0370-2693(86)90375-8
%570 citations counted in INSPIRE as of 30 Dec 2022










	\bibitem{ggkpiia}
	U.~Gran, J.~Gutowski, U.~Kayani and G.~Papadopoulos,
	``Dynamical symmetry enhancement near IIA horizons,''
	JHEP \textbf{06} (2015), 139
	doi:10.1007/JHEP06(2015)139
	[arXiv:1409.6303 [hep-th]].

	\bibitem{bgpiia}
	S.~Beck, J.~B.~Gutowski and G.~Papadopoulos,
	``Supersymmetry of IIA warped flux AdS and flat backgrounds,''
	JHEP \textbf{09} (2015), 135
	doi:10.1007/JHEP09(2015)135
	[arXiv:1501.07620 [hep-th]].

%\cite{Papadopoulos:2022ppo}

%1 citations counted in INSPIRE as of 29 Dec 2022

%\cite{Gran:2016zxk}
\bibitem{adsdes}
U.~Gran, J.~B.~Gutowski and G.~Papadopoulos,
``On supersymmetric Anti-de-Sitter, de-Sitter and Minkowski flux backgrounds,''
Class. Quant. Grav. \textbf{35} (2018) no.6, 065016
doi:10.1088/1361-6382/aaac8c
[arXiv:1607.00191 [hep-th]].
%20 citations counted in INSPIRE as of 30 Dec 2022



\bibitem{FR}
 P.~G.~O.~Freund and M.~A.~Rubin,
  {\textit{Dynamics of Dimensional Reduction,
  Phys.\ Lett.\ }} B {\bf 97} (1980) 233.



\bibitem{jose}
  B.~S.~Acharya, J.~M.~Figueroa-O'Farrill, C.~M.~Hull and B.~J.~Spence,
  {\textit{Branes at conical singularities and holography,
  Adv.\ Theor.\ Math.\ Phys.\ }}  {\bf 2} (1999) 1249
  [hep-th/9808014].

  \bibitem{cvetic}
  M.~Cvetic, H.~Lu, C.~N.~Pope and J.~F.~Vazquez-Poritz,
  {\textit{AdS in warped space-times,
  Phys.\ Rev.\ D}} {\bf 62} (2000) 122003
  [hep-th/0005246].

  \bibitem{lust1}
  D.~Lust and D.~Tsimpis,
  {\textit{Supersymmetric AdS(4) compactifications of IIA supergravity,
  JHEP}} {\bf 0502} (2005) 027
  [hep-th/0412250].

\bibitem{lust2}
   D.~Lust and D.~Tsimpis,
  {\textit{New supersymmetric AdS(4) type II vacua,
  JHEP}} {\bf 0909} (2009) 098
  [arXiv:0906.2561 [hep-th]].

  \bibitem{passias1}
  F.~Apruzzi, M.~Fazzi, A.~Passias, D.~Rosa and A.~Tomasiello,
  {\textit{$AdS_6$ solutions of type II supergravity,
JHEP}} {\bf 1411} (2014) 099 [{\textit{Erratum ibid JHEP}} {\bf 1505} (2015) 012]
  [arXiv:1406.0852 [hep-th]].

\bibitem{rosa}
   F.~Apruzzi, M.~Fazzi, D.~Rosa and A.~Tomasiello,
  {\textit{All $AdS_7$ solutions of type II supergravity,
  JHEP}} {\bf 1404} (2014) 064
  [arXiv:1309.2949 [hep-th]].

  %\cite{Gutowski:2017edr}
\bibitem{jggpx}
J.~Gutowski and G.~Papadopoulos,
``On supersymmetric AdS$_{6}$ solutions in 10 and 11 dimensions,''
JHEP \textbf{12} (2017), 009
doi:10.1007/JHEP12(2017)009
[arXiv:1702.06048 [hep-th]].




\bibitem{gppt}
G.~Papadopoulos and P.~K.~Townsend,
``Intersecting M-branes,''
Phys. Lett. B \textbf{380} (1996), 273-279
doi:10.1016/0370-2693(96)00506-0
[arXiv:hep-th/9603087 [hep-th]].

%\cite{Tseytlin:1996bh}
\bibitem{aat}
A.~A.~Tseytlin,
``Harmonic superpositions of M-branes,''
Nucl. Phys. B \textbf{475} (1996), 149-163
doi:10.1016/0550-3213(96)00328-8
[arXiv:hep-th/9604035 [hep-th]].

%\cite{Gauntlett:1996pb}
\bibitem{jgkt}
J.~P.~Gauntlett, D.~A.~Kastor and J.~H.~Traschen,
``Overlapping branes in M theory,''
Nucl. Phys. B \textbf{478} (1996), 544-560
doi:10.1016/0550-3213(96)00423-3
[arXiv:hep-th/9604179 [hep-th]].
%265 citations counted in INSPIRE as of 30 Dec 2022


%\cite{bps}
\bibitem{bps}
H.~J.~Boonstra, B.~Peeters and K.~Skenderis,
``Brane intersections, anti-de Sitter space-times and dual superconformal theories,''
Nucl. Phys. B \textbf{533} (1998), 127-162
doi:10.1016/S0550-3213(98)00512-4
[arXiv:hep-th/9803231 [hep-th]].





%\cite{Strominger:1996sh}
\bibitem{strominger}
A.~Strominger and C.~Vafa,
``Microscopic origin of the Bekenstein-Hawking entropy,''
Phys. Lett. B \textbf{379} (1996), 99-104
doi:10.1016/0370-2693(96)00345-0
[arXiv:hep-th/9601029 [hep-th]].
%3132 citations counted in INSPIRE as of 30 Dec 2022

%\cite{Callan:1996dv}
\bibitem{callan}
C.~G.~Callan and J.~M.~Maldacena,
``D-brane approach to black hole quantum mechanics,''
Nucl. Phys. B \textbf{472} (1996), 591-610
doi:10.1016/0550-3213(96)00225-8
[arXiv:hep-th/9602043 [hep-th]].
%752 citations counted in INSPIRE as of 30 Dec 2022


%\cite{Coles:1990hr}
\bibitem{coles}
R.~A.~Coles and G.~Papadopoulos,
``The Geometry of the one-dimensional supersymmetric nonlinear sigma models,''
Class. Quant. Grav. \textbf{7} (1990), 427-438
doi:10.1088/0264-9381/7/3/016
%96 citations counted in INSPIRE as of 30 Dec 2022


	

\end{thebibliography}
\end{document}